\documentclass[aps,prx,superscriptaddress,twocolumn,eprint]{revtex4-2}
\usepackage{amsmath}
\usepackage{amsthm}
\usepackage{mathtools}
\usepackage{hyperref}
\usepackage{amssymb}
\usepackage{physics}
\usepackage{graphicx}
\graphicspath{ {figures/} }
\usepackage[caption=false]{subfig}
\usepackage{bm}
\usepackage[dvipsnames]{xcolor}
\usepackage{float}

\usepackage{setspace} 

\usepackage[section]{placeins} 

\usepackage{comment}

\usepackage{empheq}

\usepackage{tikz}
\usetikzlibrary{positioning, calc}

\newcommand{\p}[2]{\ensuremath{\frac{\partial #1}{\partial #2}}} 

\newcommand{\beq}{\begin{equation}}
\newcommand{\eeq}{\end{equation}}

\begin{document}

	\title{Active topological defect absorption by a curvature singularity}
	\author{Farzan Vafa}
	\affiliation{Center of Mathematical Sciences and Applications, Harvard University, Cambridge, MA 02138, USA}
	\author{David R. Nelson}
	\affiliation{Department of Physics, Harvard University, Cambridge, MA 02138, USA}
	\author{Amin Doostmohammadi}
	\affiliation{Niels Bohr Institute, University of Copenhagen, Blegdamsvej 17, Copenhagen 2100, Denmark}
	\date{\today}
	
	\begin{abstract}
		
	Using the Born-Oppenheimer approximation, we present a general description of topological defects dynamics in $p$-atic materials on curved surfaces, and simplify it in the case of active nematics. We find that activity induces a geometric contribution to the motility of the $+1/2$ defect. Moreover, in the case of a cone, the simplest example of a geometry with curvature singularity, we find that the motility depends on the deficit angle of the cone and changes sign when the deficit angle is bigger than $\pi$, leading to the change in active behavior from contractile (extensile) to extensile (contractile) behavior. Using our analytical framework, we then identify for positively charged defects the basin of attraction to the cone apex and present closed-form predictions for defect trajectories near the apex. The analytical results are quantitatively corroborated against full numerical simulations. Provided the capture radius is small compared to the cone size, the agreement is excellent.

	\end{abstract}
	
	\maketitle
	
	\tableofcontents

	\section{Introduction}

	Active nematics consist of elongated apolar (head-tail symmetric on average) units that extract energy from their surrounding to generate active forces, leading to self-sustained flows~\cite{simha2002hydrodynamic,doostmohammadi2018active}. 
	Nematic order has been widely reported in biological systems, ranging from subcellular filaments~\cite{sanchez2012spontaneous,keber2014topology,zhang2018interplay}, to bacterial biofilms~\cite{dell2018growing,you2018geometry,copenhagen2021topological} and cell monolayers~\cite{duclos2018spontaneous,blanch2018turbulent}. In two-dimensional nematics, because of the head-tail symmetry, the lowest energy defects are $\pm 1/2$ disclinations~\cite{gennes1993the}, around which the order parameter field rotates by $\pm \pi$. Both $+1/2$ comet-shaped and $-1/2$ trefoil-shaped topological defects have recently been found to be central in many biological
	functions, e.g., cell extrusion and apoptosis in mammalian epithelia~\cite{saw2017topological},
	neural mound formation~\cite{kawaguchi2017topological}, bacterial competition~\cite{meacock2021bacteria}, and limb origination in the
	simple animal \emph{Hydra}~\cite{maroudas2020topological} (see~\cite{doostmohammadi2021physics, shankar2022topological, bowick2022symmetry} for recent reviews of the significance of topological defects in biological systems).
	
	In contrast to its passive counterpart, activity renders a comet-shaped $+1/2$ defect motile, driving it to self-propel along its axis of symmetry, in a direction dictated by the sign of the active stress: extensile (contractile) active stress drives the defect to move towards the head (tail) of the comet~\cite{doostmohammadi2018active}. A trefoil-shaped $-1/2$ defect, on the other hand, due to its three-fold rotational symmetry does not become motile, unless external forces or boundary conditions violate the three-fold symmetry.
	 
	Previous theoretical work has described multi-defect active nematics by treating the topological defects as quasiparticles, with elastic distortions of the nematic texture and active flows mediating the effective interactions in phenomenological models~\cite{keber2014topology, giomi2013defect, shankar2018defect}. Recently, a systematic formulation based on the Born-Oppenheimer approximation was employed for active nematics~\cite{zhang2020dynamics, vafa2020multi-defect, vafa2022defectDynamics} by treating the defect positions as slow degrees of freedom and the nematic texture as a fast degree of freedom that instantaneously responds to the slow motion of the defects.
	
	The discussion thus far has focused on flat surfaces; however, much of biology takes stage on curved surfaces. Specifically, various natural and synthetic developmental processes involve formation of three-dimensional curved structures from cell monolayers. Striking examples include epithelial dome formation~\cite{latorre2018active}, embryo gastrulation~\cite{behrndt2012forces}, and in-vitro tissue regeneration~\cite{matejvcic2022mechanobiological}. In addition, more recently, nematic organization of active entities on flat surfaces has been linked to three-dimensionalization processes and in particular to the formation of protrusions in confined myoblast cells~\cite{guillamat2022integer}, from cytoskeleton of animal {\it Hydra} during morphogenesis~\cite{maroudas2020topological}, and from the membrane of eukaryotic cells in the form of cellular fingers, known as filopodia~\cite{leijnse2022filopodia}. 
 
 Some progress has been made on curved surfaces in the {\em passive} context. Ref.~\cite{lubensky1992orientational} analytically derived the ground state defect configurations on a sphere and torus, and the interaction between liquid crystalline order and curved substrates was studied in~\cite{park1996topological,vitelli2004anomalous}, where it was shown that curvature gives rise to an effective topological charge density. For a cone, this corresponds to negative topological charge concentrated at the apex, and a simple argument was recently presented in~\cite{zhang2022fractional} for the case of free boundary conditions. Ref.~\cite{vafa2022defectAbsorption} re-derived the induced charge result of Vitelli and Turner~\cite{vitelli2004anomalous} and in the context of a cone with tangential boundary conditions used it to determine the ground state defect configuration given a fixed number of topological defects. As a cone, which has a curvature delta function singularity at the apex, is the simplest example of nontrivial curved geometry, we continue to study the dynamics of {\em active} nematics on a cone geometry here.
	
	This paper is organized as follows. We begin in Sec.~\ref{sec:formulation} by formulating a minimal model of a general $p$-atic texture on an arbitrarily curved surface. Working deep in the ordered limit, and taking advantage of isothermal coordinates (recently introduced in the context of liquid crystals~\cite{vafa2022active,vafa2022defectAbsorption}), we write down explicitly the quasistatic multi-defect solution in the passive setting on a curved surface. Then in Sec.~\ref{sec:passive} we study the dynamics of the texture by working within the Born-Oppenheimer approximation: we assume that the defects move slowly (valid in the limit of low defect density and low activity) and that the nematic texture instantaneously readjusts itself in response. The partial differential equation for the nematic texture is thus reduced to a set of ordinary differential equations for the effective defect positions. Upon introducing activity in Sec.~\ref{sec:active}, we first derive the hydrodynamic equations of a compressible active nematic film on a curved surface. Working within the Born-Oppenheimer approximation (as in the passive case) yields a number of new results. In particular, applying the framework to the case of active nematics we show that the motility of a $+1/2$ defect picks up an active geometric contribution, and changes sign as the curvature is increased. For a $+1/2$ defect on a cone, we identify the basin of attraction to the apex and predict its trajectory. Throughout the paper, we quantitatively check our theoretical predictions against full numerical simulations of active nematics on a cone, explore the effect of boundary conditions, and scrutinize the validity of the Born-Oppenheimer approximation, finding excellent agreements. Sec.~\ref{sec:discussion} summarizes our main results and comments on applications. Most of the technical details are relegated to Appendices \ref{app:derivation}-\ref{app:trajectory}.

	\section{Formulation of a minimal model in isothermal coordinates}
    \label{sec:formulation}
Here we review the framework for describing a $p$-atic texture on a curved surface, following the presentation in Ref.~\cite{vafa2022defectAbsorption}. The case of a nematic texture is recovered by simply setting $p=2$. 

\subsection{Metric}
In two dimensions it is always possible to choose local complex coordinates $z$ and $\bar z$, known as isothermal (or conformal) coordinates, such that the length of the interval squared can be written as~\cite{gauss1822on},
	\beq ds^2 = g_{z\bar z} dz d\bar z + g_{\bar z z} d\bar z dz = 2g_{z\bar z}|dz|^2 = e^{\varphi}|dz|^2, \eeq
    where $e^{\varphi}$ is the conformal factor that describes position-dependent isotropic stretching, and $g_{z\bar{z}} = e^{\varphi}/2$ is the metric. In Cartesian coordinates,
    \beq z = x + iy, \qquad \bar z = x - iy, \eeq
    from which follows that
    \beq ds^2 = e^{\varphi(x,y)}(dx^2 + dy^2). \eeq

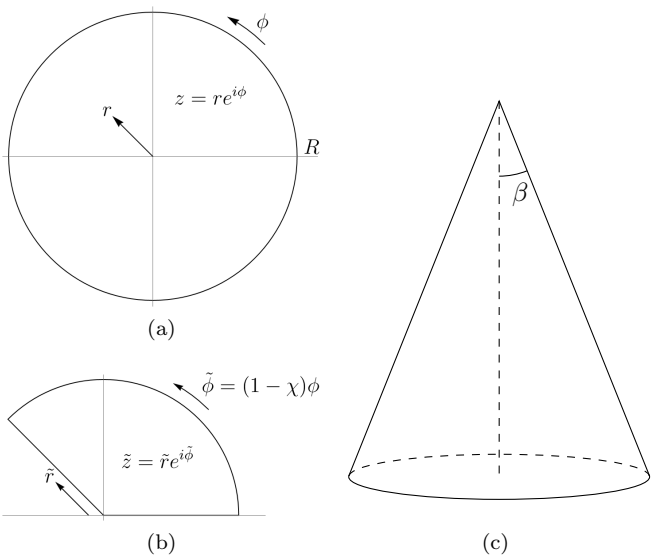
\begin{figure}[t]
\begin{minipage}[b]{.49\columnwidth}
        \centering
            \subfloat[]{\includegraphics[width=\textwidth]{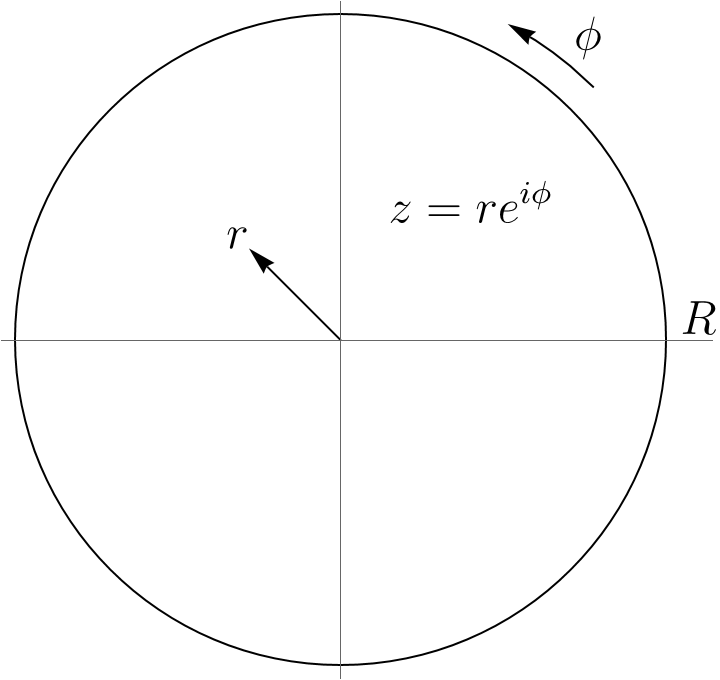}}\\
        \subfloat[]{\includegraphics[width=\textwidth]{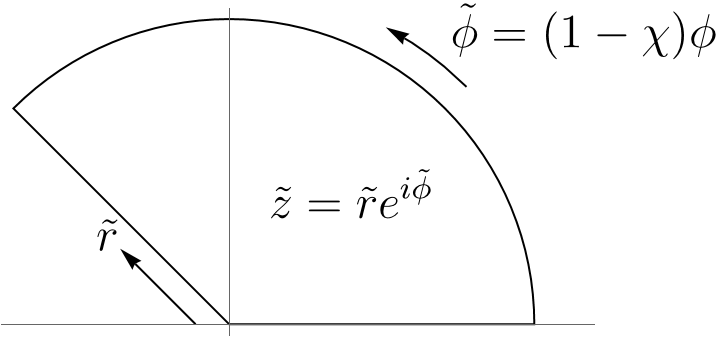}}
    \end{minipage}
    \hfill
    \begin{minipage}[b]{.49\columnwidth}
        \centering

\subfloat[]{\begin{tikzpicture}
	
	\newcommand{\radiusx}{2}
	\newcommand{\radiusy}{.3}
	\newcommand{\height}{5}
	
	\coordinate (a) at (-{\radiusx*sqrt(1-(\radiusy/\height)*(\radiusy/\height))},{\radiusy*(\radiusy/\height)});
	
	\coordinate (b) at ({\radiusx*sqrt(1-(\radiusy/\height)*(\radiusy/\height))},{\radiusy*(\radiusy/\height)});
	
	\draw (a)--(0,\height)--(b);
	
	\draw[dashed] (0,\height)--(0,0);

	\begin{scope}
		\clip ([xshift=-2mm]a) rectangle ($(b)+(1mm,-2*\radiusy)$);
		\draw circle (\radiusx{} and \radiusy);
	\end{scope}
	
	\begin{scope}
		\clip ([xshift=-2mm]a) rectangle ($(b)+(1mm,2*\radiusy)$);
		\draw[dashed] circle (\radiusx{} and \radiusy);
	\end{scope}
	
	\draw (0,\height-1) arc (-90:-90+atan(\radiusx/\height):1) node[below, pos=.75]{$\beta$}; 
\end{tikzpicture}}
    \end{minipage}  
\caption{Schematic of coordinate systems for a cone: (a) Isothermal coordinates $z = re^{i\phi}$. (b) Physical coordinates $\tilde z = \tilde r e^{i\tilde\phi}$, corresponding to an unrolled cone with a missing sector, with angle fraction $\chi$. (c) A diagram of the cone in 3D with cone half angle $\beta$, where $\sin\beta = 1-\chi$.}
\label{fig:cone}
\end{figure}

    \noindent{\bf Conical geometry.}
    The conical geometry is a prime example that we study in this paper. For a cone with half angle $\beta$, the metric can be obtained from $g_{z\bar{z}}=e^{\varphi}$ with
	\beq \varphi = -\chi \ln z \bar z, \label{eq:cone} \eeq
	where $\chi=1-\sin\beta$. As such, $2\pi\chi$ describes the deficit angle of the cone, and for example $\chi = 0$ corresponds to a disk (no fraction missing). Another set of useful coordinates are physical coordinates $\tilde z$, which correspond to unrolling a cone to form a planar disk with a missing sector, are related to $z$ via the coordinates 
 \beq \tilde z= \frac{z^{1-\chi}}{1-\chi}. \label{eq:ztilde}\eeq
 In these coordinates,
 \beq ds^2 = d\tilde z^2, \eeq
 which is flat everywhere except at the origin. The distances from the apex in these two coordinate systems are related by
 \beq \tilde r = \frac{1}{1-\chi} r^{1-\chi} . \eeq
 At the origin, $\tilde z$ is not defined and there is a conical singularity with deficit angle $2\pi\chi$. See Fig.~\ref{fig:cone} for a schematic of various coordinate systems for a cone.

~\\

\subsection{Tensors and covariant derivatives}

    Let $\mathbf{T}$ denote a traceless real symmetrized rank-$p$ tensor that describes the $p$-atic order. Then in isothermal coordinates, since $\mathbf{T}$ is traceless (contraction of any pair of indices vanishes), $\mathbf{T}$ has only two non-zero components $T \equiv T^{z\ldots z}$ and $\bar T \equiv T^{\bar z \ldots \bar z}$, where here ellipses denote $p$ copies. Also, by reality, $T = (\bar T)^*$.
    
    For ease of notation, let $\nabla \equiv \nabla_z$ and $\bar\nabla \equiv \nabla_{\bar z}$ denote the covariant derivatives with respect to $z$ and $\bar z$, respectively. Covariant derivatives of $\mathbf{T}$ are quite simple using isothermal coordinates
	\begin{subequations}
 \begin{gather}
    \nabla T = \partial T + p (\partial \varphi) T, \qquad \bar \nabla T = \bar \partial T, \label{eq:nabla1}\\ 
	\bar\nabla \bar T = \bar \partial \bar T + p (\bar\partial \varphi) \bar T, \qquad \nabla \bar T = \partial \bar T, \label{eq:nabla2}
  \end{gather}
  \label{eq:nabla}
\end{subequations}
where $\partial \equiv \partial/\partial z$ and $\bar{\partial} \equiv \partial / \partial\bar z$. Note the asymmetry between the first and second equations of Eq.\eqref{eq:nabla1} and Eq.~\eqref{eq:nabla2}. In Cartesian coordinates, we have
 \beq \p{}{z} = \frac{1}{2}\left(\p{}{x} - i \p{}{y}\right), \qquad \p{}{\bar z} = \frac{1}{2}\left(\p{}{x} + i \p{}{y}\right),\label{eq:partial}\eeq
and  
\beq \nabla_z = \frac{1}{2}\left(\nabla_x - i \nabla_y\right), \qquad \nabla_{\bar z} = \frac{1}{2}\left(\nabla_x + i \nabla_y\right). \eeq
See Table~\ref{tab:isotherm} for the correspondence between isothermal coordinates and Cartesian coordinates.

\begin{table*}[t]
    \centering
    \begin{tabular}{c|c|c}
        symbol in isothermal coordinates & physical interpretation & Cartesian representation \\
        \hline
        \hline
        ($z$, $\bar{z}$) & coordinates & ($x+i y$, $x-i y$)\\
        \hline
        $\partial \equiv \partial/\partial_z$ & partial derivative & $\frac{1}{2}\left(\frac{\partial}{\partial x} - i \frac{\partial}{\partial y} \right)$\\
        \hline
        $\bar{\partial} \equiv \partial/\partial_{\bar{z}}$ & partial derivative & $\frac{1}{2}\left(\frac{\partial}{\partial x} + i \frac{\partial}{\partial y} \right)$\\
        \hline
        $\nabla \equiv \nabla_z$ & covariant derivative & $\frac{1}{2}\left(\nabla_x - i \nabla_y \right)$\\
        \hline
        $\bar{\nabla} \equiv \nabla_{\bar{z}}$ & covariant derivative & $\frac{1}{2}\left(\nabla_x + i \nabla_y \right)$\\
        \hline
        $T = T^{\overbrace{z...z}^{p-{\text{times}}}},~\bar{T} = T^{\overbrace{\bar{z}...\bar{z}}^{p-{\text{times}}}}$ & the only two non-zero components of $\mathbf T$ & --\\
        \hline
        $Q = Q^{zz}$, & the only two non-zero components of & $Q=(Q^{xx}-Q^{yy})+i(2Q^{xy})$,\\
        $\bar{Q} = Q^{\bar{z}\bar{z}}$ & the nematic tensor $\mathbf Q$ & $\bar{Q}=(Q^{xx}-Q^{yy})-i(2Q^{xy})$\\
        \hline
    \end{tabular}
    \caption{Definitions and correspondence between isothermal coordinates and Cartesian coordinates.}
    \label{tab:isotherm}
\end{table*}

\subsection{Free energy}

 The free energy $\mathcal F$ (in a one Frank constant approximation for $p=2$) in terms of expansion in powers of the order parameter $T$ and its gradients can be written as
	\beq \mathcal{F} = \int d^2z \sqrt{g}[K |\nabla T|^2 + K' |\bar \nabla T|^2 + \epsilon^{-2} (1 - S_0|T|^2)^2],\label{eq:minimal}\eeq
 where explicitly
  \begin{align}
   |\nabla T|^2 &= g_{z \bar z}^{p-1}\nabla T \bar\nabla \bar T, \qquad |\bar \nabla T|^2 = g_{z \bar z}^{p-1}\bar\nabla T \nabla \bar T \\
   &\qquad \qquad |T|^2 = g_{z \bar z}^p T \bar T .  
 \end{align} 
	Here $K,K' >0$ are Frank elastic type terms~\cite{frank1958liquid,berreman1984tensor}, and in regions of zero Gaussian curvature (such as any point on a cone other than the apex), the two terms are equivalent by integration by parts. The last term is the Landau-De Gennes type free energy that governs the isotropic-$p$-atic transition, with $\epsilon$ controlling the microscopic coherence length and $S_0$ sets the equilibrium magnitude of the $p$-atic order. Without loss of generality we set $S_0 = 2^p$.
	
	On using the free energy in Eq.~\eqref{eq:minimal}, the $p$-atic order parameter $T$ can be determined by minimizing the free energy. Deep in the ordered limit ($\epsilon \ll 1$), we have
	\beq 2^p|T|^2 = 1. \eeq
	Writing the order parameter in terms of its amplitude $\mathcal{A}$ and phase $\alpha$, we find $T^{z\ldots z} = \mathcal{A}^{z\ldots z}e^{i\alpha} = \mathcal{A}e^{i\alpha}$, which leads to
	\beq T = \mathcal{A}e^{i\alpha} = (2g_{z\bar z})^{-p}e^{i\alpha} = e^{-\frac{p}{2}\varphi + i\alpha}.\label{eq:A}\eeq
	In other words, the contribution of the potential term to the free energy vanishes, and the free energy simplifies to~\footnote{Note that this expression is valid far from defects and boundaries.}
	\beq \mathcal F = 2^{p-1}\int d^2z \sqrt{g}[K|\nabla T|^2 + K'|\bar\nabla T|^2] \;\label{eq:minimal2} . \eeq
	
	Upon substitution of Eq.~\eqref{eq:A} into Eq.~\eqref{eq:minimal2}, the free energy can be written in terms of $\varphi$ (the log of the conformal factor) and the order parameter's phase $\alpha$ as
	\beq	\mathcal F = (K+K')\int d^2z\left|\left(\frac{p}{2}\right)\partial\varphi + i\partial\alpha\right|^2, \label{eq:FSimple}\eeq
	where we have used
    \begin{subequations}
	\begin{align}
		\nabla T &= \left(\frac{p}{2}\partial\varphi + i\partial\alpha\right)T,\\
		\bar\nabla T &= \left(-\frac{p}{2}\bar\partial\varphi + i\bar\partial\alpha\right)T.
	\end{align}
    \end{subequations}
    Minimizing $\mathcal F$ (Eq.~\eqref{eq:FSimple}) with respect to the order parameter phase $\alpha$ gives
    \beq \partial \bar \partial \alpha = 0 \label{eq:LaplaceEq} .\eeq
    In the presence of a topological defect of charge $\sigma \in \mathbb{Z}/p$, the phase $\alpha$ will wind by $2\pi p \sigma$. Thus a solution to Eq.~\eqref{eq:LaplaceEq} with a single defect $j$ at $z_j$ with charge $\sigma_j$ is (neglecting image charges that allow us to impose various boundary conditions)
    \beq \alpha_0 = -\frac{i}{2} (p\sigma_j) \ln \frac{z - z_j}{\bar z - \overline{z_j}} , \label{eq:theta_0}\eeq
  which results in the energy of a single $p$-atic topological defect of charge $\sigma_j$~\cite{vafa2022defectAbsorption}
	\beq \mathcal F_0 = -\frac{\pi p^2}{2} (K+K') \left(\sigma_j - \frac{1}{2}\sigma_j^2\right)\varphi(z_j). \label{eq:F}\eeq
    Note that $\varphi({z_j})$ is a purely geometrical contribution that is set by the shape of the curved space on which the $p$-atic texture resides. Recall from Eq.~\eqref{eq:cone} that, for a cone with half-angle $\beta$, we have
    \beq \varphi(z_zj) = - \chi \ln (z_j \bar z_j) = -(1 - \sin\beta)\ln(z_j \bar z_j) . \eeq
    
     Having set up the isothermal coordinates and having defined the energy of topological defects for $p$-atics on arbitrary geometries, we now turn to defect dynamics. Our goal is to establish an effective description of topological defects dynamics through a set of ODEs that describe Langevin-type behavior for the positions of topological defects that behave like charged quasi-particles, where the liquid crystal textures relax instantaneously. We begin with the passive case, and then extend to active defect dynamics.
     
	\section{Equations of defect dynamics: passive case} \label{sec:passive}

    We begin by considering the passive case, while keeping the description general for $p$-atic on an arbitrary geometry with conformal factor $e^{\varphi(z, \bar z)}$. We obtain $p$-atic dynamics in the limit of low defect density by the following procedure: we minimize the mean squared deviation of the dynamics on the inertial manifold $T_0$ from the exact equation of motion. Specifically, we assume that the defects move slowly and that in response the $p$-atic texture instantaneously readjusts itself. When studying the quantum mechanics of light-weight electrons bonding atoms with much heavier nuclei, this is known as the Born-Oppenheimer approximation~\cite{landau2013quantum}. Similar approximations were made to study the dynamics of vortices in superfluid helium films driven out of equilibrium by an oscillating substrate~\cite{ambegaokar1980dynamics} and the dislocation-mediated elongation of the peptidoglycan  cell walls of bacteria~\cite{amir2012dislocation}.

     \subsection{Born-Oppenheimer approximation}

    In the passive case, and in the absence of active flows, we assume the exact equation of the motion for the defect phase $\alpha$ is controlled by the relaxational dynamics
    \beq \partial_t \alpha ( z,{\bar z},t) = 
	- \gamma^{-1}\frac{1}{\sqrt{g}}\frac{\delta \mathcal F}{\delta \alpha},\label{eq:complexQ}\eeq
    where $\gamma$ is the rotational diffusion coefficient and the dependence on geometry is manifest through the metric $g$. We take as our Born-Oppenheimer ansatz~\cite{vafa2020multi-defect,vafa2022defectDynamics},
    \beq \alpha(z,\bar z,t) = \alpha_0(z,\bar z | \{z_i(t)\}),\eeq
    where $z_i(t)$ are time-dependent positions for defects, and find $z_i(t)$ by minimizing the mean squared deviation between the dynamics on the inertial manifold and the exact equation of motion (Eq.~\eqref{eq:complexQ}),
	\begin{align}
		E &= \int d^2z \sqrt{g}\left| \frac{d}{dt} \alpha_0( z,{\bar z}|\{ z_i (t) \}) + \gamma^{-1}\frac{1}{\sqrt{g}}\frac{\delta \mathcal F[\alpha_0]}{\delta \alpha}\right|^2 \nonumber\\
		&= \int d^2z \sqrt{g} \left| \dot z_i\partial_i \alpha_0 + \dot{\bar z}_i \bar\partial_i \alpha_0 + \gamma^{-1}\frac{1}{\sqrt{g}}\frac{\delta \mathcal F[\alpha_0]}{\delta \alpha}\right|^2,
		\label{nematic_eq:E}
	\end{align} 
	with respect to $\dot z_i$, where $\alpha_0$ is given by Eq.~\eqref{eq:theta_0}. Doing so~\cite{vafa2020multi-defect} leads to the following coupled set of simple ODEs for the defect dynamics in isothermal coordinates
	\beq \mathcal M_{ij}\dot z_j  + \mathcal N_{ij}\dot{\bar z}_j   = -\p{{\mathcal F}_0}{\bar z_i}\;,\label{nematic_defect dynamics}\eeq
	where $-\p{{\mathcal F}_0}{\bar z_i}$ is the Coulombic force computed by differentiating the Coulombic potential (Eq.~\eqref{eq:F}). (Were we to include thermal fluctuations, Langevin noise sources would appear in Eq.~\eqref{nematic_defect dynamics}.) The tensors $\mathcal M_{ij}$ and $\mathcal N_{ij}$ are collective mobility matrices that describe the effective response of a topological defect to the Coulombic forces, and are geometry dependent
    \begin{subequations}
	\begin{align}
		\mathcal M_{ij} &= \int d^2z \sqrt{g}\bar\partial_i \alpha_0 \partial_j \alpha_0\label{nematic_eq:M},\\
		\mathcal N_{ij} &= \int d^2z \sqrt{g}\bar\partial_i \alpha_0 \bar \partial_j \alpha_0,\label{nematic_eq:N}
	\end{align}
    \end{subequations}
	where $g$ is the metric. Taken together, Eqs.~\eqref{nematic_defect dynamics}-\eqref{nematic_eq:N} along with Eq.~\eqref{eq:theta_0} and Eq.~\eqref{eq:F} for $\alpha_0$ and $\mathcal{F}_0$, respectively, describe the dynamics of passive topological defects of charge $\sigma_i$ located at $z_i$ for a $p$-atic texture on a curved geometry that is described by the metric $g_{z\bar{z}}=e^{\varphi}/2$. We next apply this framework to first describe defect dynamics on a cone geometry, and will then extend the framework to investigate the effect of activity-induced flows.

\subsection{Evaluation of dynamical equation}
We first compute the collective mobility $\mathcal M_{ii}$ and $\mathcal N_{ii}$ for a single defect of charge $\sigma_i$ located at $z_i$ on a cone, as the simplest example of nontrivial curved geometry. Up to now, our discussion has been general. Here we will continue to be general (for example, write in terms of $p$ and generic defect charge $\sigma_i$), but for simplicity we consider the geometry of a cone with cone apex at the origin, described by $\varphi = -\ln(z\bar z)$ and the cone radius $R \gg a^{1/(1-\chi)}$, where $a$ is the defect core size. We assume that the defect is sufficiently far from the cone base so we can ignore any image charges, despite the long-range interactions. The computations of $\mathcal M_{ii}$ and $\mathcal N_{ii}$ are presented in Appendix~\ref{app:mobilities}, and since $\mathcal N_{ii}$ is subleading to $\mathcal M_{ii}$, we shall neglect $\mathcal N_{ii}$.
    
    When the defect is sufficiently far from the apex, i.e., $r_i = |z_i| \gg a^{1/(1-\chi)}$, $\mathcal M_{ii}$ is given by
    \beq \mathcal M_{ii} = \frac{\pi}{4} (p\sigma_i)^2 r_i^{-2\chi}\ln(r_i^{(1-\chi)}/a),\label{eq:Mcone1}  \eeq
    and thus
    \beq  \dot z_i = -3\gamma^{-1} (K+K')\chi \frac{1}{(p\sigma_i)^2\ln(r_i^{(1-\chi)}/a)}\frac{r_i^{2\chi}}{\overline{z_i}}.\eeq
 	When the defect is close to the apex, instead we find
	\beq \mathcal M_{ii} =  \frac{\pi}{4}(p\sigma_i)^2\frac{1}{2\chi} \left(\delta^{-2\chi} - R^{-2\chi} \right) , \label{eq:Mcone2} \eeq
 and thus
 \beq  \dot z_i = -3\gamma^{-1} (K+K')\chi  \frac{1}{(p\sigma_i)^2\left(\delta^{-2\chi} - R^{-2\chi} \right) }\frac{r_i^{2\chi}}{\overline{z_i}},\eeq
	where $\delta = a r_i^{-\chi} \sim a^{1/(1-\chi)}$ for $r_i \sim \delta$. Taken together, the equations of defect dynamics for a $p$-atic texture on a cone with deficit angle $2\pi \chi$ become
 \begin{equation}
\dot z_i=\frac{-3\gamma^{-1} (K+K')\chi}{(p\sigma_i)^2}\frac{r_i^{2\chi}}{\overline{z_i}}
    \begin{cases}
    \frac{1}{\ln(r_i^{(1-\chi)}/a)},& r_i \gg \delta\\
    ~\\
    2\chi\frac{r_i^{-2\chi}}{\left(\delta^{-2\chi} - R^{-2\chi} \right)}, & r_i \sim \delta
    \end{cases}
    \label{eq:dzidtPassive}
\end{equation}
 where $r_i = |z_i|$, $a \sim \delta^{1-\chi}$ is the defect core size near the apex, $R$ is the cone radius, $\gamma$ is the orientational diffusion, and $K,~K'$ denote orientational elasticities. We comment that in the case of a disk ($\chi = 0$), the mobility reduces to the well-known defect friction~\cite{denniston1996disclination}.
 ~\\

 \subsection{Global vs. local defect position}
	
	It is important to elaborate upon what we mean by the defect position $z_i$. The standard definition of the defect position is determined by the zero of the tensor order parameter $T$ where the phase of $T$ winds around. This gives a local definition of defect position which does not depend on the profile of the texture far away. However, one might instead be interested in a defect position determined by fitting the global texture to a defect profile ansatz. This global definition of course may be different from the local definition. The equations that we have derived have been obtained by fitting the global ansatz and thus the defect dynamics given by Eq.~\eqref{eq:dzidtPassive} depends on the global texture, and it not necessarily the same as the local defect position.
	
	To the extent that our ansatz for the defect profile is good locally, the two notions of local vs global defect positions should give similar results. When the two notions differ substantially, the Born-Oppenheimer ansatz breaks down for the local description of the texture. We will check when these two notions agree and disagree. 

\subsection{Comparison with simulations}
  
  In order to check our analytical formulation, we compare the dynamics of a single topological defect of charge $\sigma$ calculated from Eq.~\eqref{nematic_defect dynamics} using the mobility $M_{ii}$ from Eq.~\eqref{eq:Mcone2}, with full numerical simulations of the passive nematic, i.e. $p=2$, texture on a cone.

    Although we wrote our equations in terms of the phase $\alpha$ of the order parameter, in our simulations we numerically evolve the full nematic texture ${\bf T} = {\bf Q}$ ($p=2$) itself explicitly, according to
    \begin{align}
    \gamma\partial_t Q &= -\frac{1}{\sqrt{g}} g^{z\bar z} g^{z \bar z } \frac{\delta \mathcal F}{\delta \bar Q} \nonumber \\
    &= g^{z\bar z} \left(K\nabla \bar\nabla + K' \bar \nabla \nabla\right) Q + 2\epsilon^{-2}S_0(1 - S_0|Q|^2)Q,\label{eq:num}
    \end{align}
    where $\mathcal F$ is given in Eq.~\eqref{eq:minimal}, $S_0 = 2^p$, and $Q$ is the nematic tensor, which in isothermal coordinates can be written in terms of its Cartesian components as
    \beq Q = (Q^{xx} - Q^{yy}) + 2i Q^{xy}.\label{Qzz}\eeq
    Without loss of generality, since a cone has zero Gaussian curvature everywhere except at the apex, we set $K'=0$ in the simulations. We choose periodic boundary conditions on a square box, the simplest boundary conditions to simulate. We expect the choice of boundary conditions to not be important for trajectories and times such that the defect distance $r(t)$ to the apex is much smaller than the size of the box $R$, i.e. $r(t) \ll R$. We also checked this expectation by checking that defect trajectories (discussed later in the paper) don't change if we change the boundary conditions to free~\cite{zhang2022fractional} or tangential boundary conditions~\cite{vafa2022defectAbsorption} at the boundary of a disk. By tangential boundary conditions, we mean that the nematic director is tangential to the boundary, and by free boundary conditions, we mean that the radial component of the gradient of $Q$ vanishes. Explicitly, we impose at the boundary of a square grid of the isothermal coordinate space of Fig.~\ref{fig:cone}(a), where the boundary is given by $z = R e^{i\phi}$ and $\phi$ is the azimuthal coordinate):
    \beq\begin{cases}
        Q = e^{-\varphi + i \phi}, & \text{tangential boundary conditions} \\
        ~\\
        \nabla_n Q = 0, & \text{free boundary conditions},
    \end{cases} \eeq
    where $\nabla_n$ is the covariant derivative in the direction normal to the boundary. In our simulations, since defects are initially placed near the apex and far from the boundary, the precise choice of boundary condition should not matter for defect trajectories that remain near the apex.
    In our simulations, we solve Eq.~\eqref{eq:num} numerically using the method of lines~\cite{schiesser2012numerical}. The temporal evolution is performed through a predictor-corrector scheme~\cite{press1992multistep} and spatial derivatives are evaluated using five-point stencil central differences.
    
	For $\chi>0$, a defect of charge $\sigma$ is attracted to the apex with geometric force proportional to $\chi(\sigma - \frac{1}{2}\sigma^2)/r$~\cite{vafa2022defectAbsorption}, which is indeed what we observe in our simulations: once a defect of charge $\sigma$ is absorbed by the apex, then the effective charge at the apex is reduced. Now the net force on any remaining charge or defect on the cone flank is proportional to $[\chi(\sigma - \frac{1}{2}\sigma^2) - \sigma \cdot \sigma]/r$. The force is repulsive for sufficiently small $\chi$, and attractive for sufficiently large $\chi$; in particular, the force vanishes at $\chi = \frac{\sigma^2}{\sigma - \frac{1}{2}\sigma^2}$. For $\sigma = 1/2$ (relevant to a $p=2$ nematic in the one Frank constant approximation), the force vanishes at $\chi = 2/3$. 
 
\begin{figure}[t]
		\centering
		\includegraphics[width=\columnwidth]{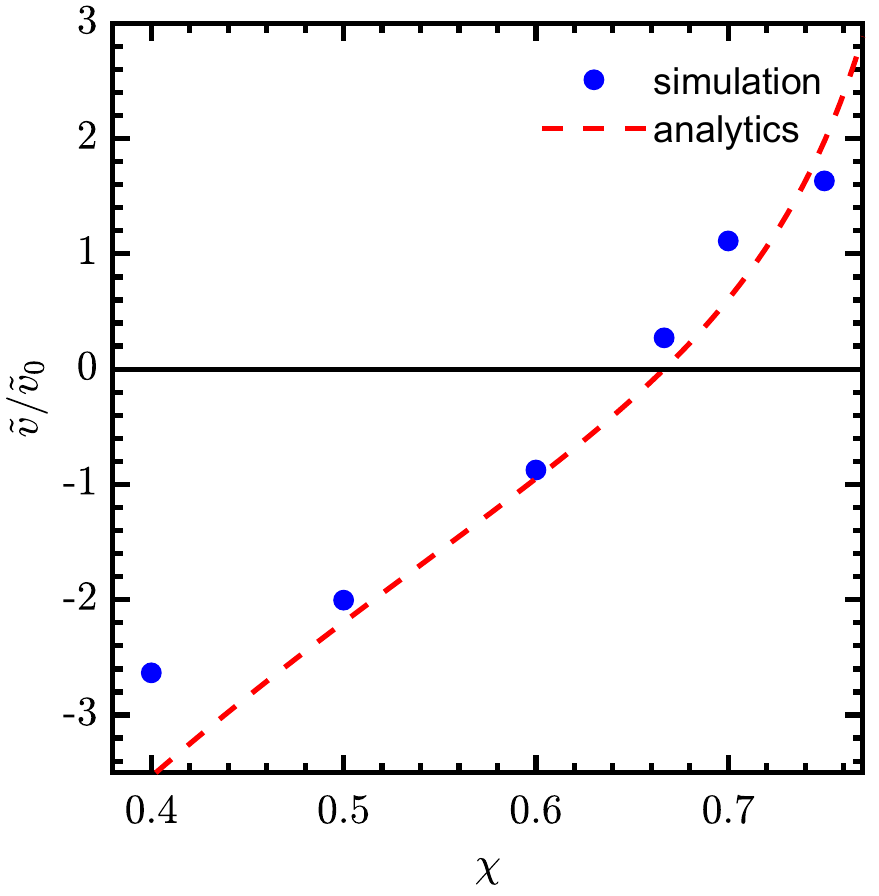}
		\caption{Initial physical global defect velocity of a $+1/2$ defect at $x/R =-0.08$ in the presence of another $+1/2$ defect already at the origin. The blue points are obtained by fitting the texture at each time to the ansatz and determining $z_i(t)$ which minimize the mean squared deviation $E$ in Eq.~\eqref{nematic_eq:E}. The red curve is fit of Eq.~\eqref{eq:dzidtTwoDefects} with $a/(\sqrt{K+K'}\epsilon) = 0.5$.
        The $y$-axis is rescaled by a passive characteristic velocity $\tilde v_0 = \gamma^{-1} (K+K')R^{2\chi-1}$.}
		\label{fig:dzidtTwoDefects}
	\end{figure}

    For a passive $+1/2$ defect at the apex, upon setting $p=2$ and $\sigma_i = 1/2$, Eq.~\eqref{eq:dzidtPassive} becomes
	\beq  \dot z_i = -\gamma^{-1} (K+K')(3\chi - 2)  \frac{1}{\ln(r_i^{(1-\chi)}/a)}\frac{r_i^{2\chi}}{\overline{z_i}}\label{eq:dzidtTwoDefects}\eeq
	We checked this formula (with $K' = 0$) using simulations (see Fig.~\ref{fig:dzidtTwoDefects}), where we find that the force does indeed change sign for $\chi \approx 2/3$.
 
	\section{Equations of defect dynamics: active case}
 \label{sec:active}
 
    So far we have limited the description to the passive dynamics of topological defects, where the motion is controlled by the relaxational dynamics of the $p$-atic texture (Eq.~\eqref{eq:complexQ}). The effect of activity enters through coupling of the relexational dynamics to the velocity field that is locally generated by the active stresses. In the overdamped limit~\cite{doostmohammadi2016stabilization,putzig2016instabilities,srivastava2016negative}, the velocity field can be directly calculated from the balance of active force and the frictional damping
    \beq \mu \vec v = \nabla \cdot \sigma^{\text{active}},\eeq
    leading to 
    \beq \vec v = \frac{1}{\mu}\nabla \cdot \sigma^{\text{active}},\label{eq:v} \eeq
    where $\mu$ is the friction coefficient and $\sigma^{\text{active}}$ denotes activity-induced stresses. For a general case of a $p$-atic, the active stress can comprise of terms proportional to the contractions of $p$-atic tensor $T$ that are allowed by symmetry~\cite{giomi2022hydrodynamic}. The active stress can therefore be considered as a function of the $p$-atic order parameter tensor $T$ as $\sigma^{\text{active}}=\sigma^{\text{active}}[T]$. It thus follows from Eq.~\eqref{eq:v} that in the general case, the activity-induced velocity field $v$ can be written in terms of the $p$-atic order parameter as $v=v[T]$. While here we keep the description general for active dynamics of any $p$-atic texture on an arbitrary geometry, in Sec.~\ref{sec:activenem} we present a detailed analyses of the dynamics for active nematics ($p=2$) on the cone and compare the theoretical predictions with full numerical simulations of the active nematics. 

\subsection{Derivation of dynamical equation}
Once the active velocity field is found from Eq.~\eqref{eq:v}, we can determine the contribution from the active velocity to the dynamics of the defect phase. To this end, Eq.~\eqref{eq:complexQ} is modified to account for balancing activity-induced flows and relaxational dynamics
    \beq \frac{D}{Dt} \alpha ( z,{\bar z}, t) = 
	- \gamma^{-1}\frac{1}{\sqrt{g}}\frac{\delta \mathcal F}{\delta \alpha},\label{eq:complexQ2}\eeq
	where $D/Dt$ is the advective derivative. Naively, one would expect
	\beq \frac{D}{Dt} \alpha = \partial_t \alpha + (v\partial  + \bar v \bar\partial)\alpha. \eeq
	However, this is not correct as $\alpha$ is anomalous: indeed, we are not advecting a conventional scalar! Under rotation of coordinates $z \to e^{i\Delta \alpha} z$, $\alpha$ transforms as $\alpha \to \alpha + p\Delta \alpha$.
	We therefore need to derive the correct form of $D \alpha/Dt $. We start from the advective derivative of a $p$-atic tensor, 
	\beq \frac{D}{Dt} T = \partial_t T + (v\nabla  + \bar v \bar\nabla)T - (\nabla v - \bar\nabla \bar v)T, \eeq
	and upon substitution of $T = e^{-\frac{p\varphi}{2} + i\alpha}$ (Eq.~\eqref{eq:A}), we find that
\begin{align}
     \frac{D}{Dt} T =  \frac{D}{Dt}\alpha (i T) = & \partial_t\alpha (i T) +   (v\partial \alpha  + \bar v \bar\partial \alpha)(i T) \nonumber \\
     &- (\partial v - \bar\partial \bar v)T \nonumber\\
     &+ (p/2 - 1)(v\partial\varphi - \bar v \bar\partial \varphi)T.
\end{align}

	The last term in the above equation arises from the geometric contributions to the advection and the co-rotation, and cancels out only for $p=2$, i.e. nematics. On dividing through by $i T$ we obtain
	\beq \frac{D}{Dt}\alpha = \partial_t \alpha + (v\partial  + \bar v \bar\partial)\alpha + i(\partial v - \bar\partial\bar v) - i (p/2 - 1)(v\partial\varphi - \bar v \bar\partial \varphi),\label{eq:Dttheta}\eeq
	where the velocity field $v$ in our overdamped limit is set by the active stresses (Eq.~\eqref{eq:v}).
 
	With Eq.~\eqref{eq:Dttheta}, we can now obtain active dynamics in the limit of weak activity and low defect density by following the same procedure as in Sec~\ref{sec:passive}: we minimize the mean squared deviation of the dynamics on the inertial manifold $T_0$ from the exact equation of motion Eq.~\eqref{eq:Dttheta}. This procedure~\cite{vafa2020multi-defect} leads to the following coupled set of ODEs for the active defect dynamics
	\beq \mathcal M_{ij}\dot z_j  + \mathcal N_{ij}\dot{\bar z}_j   = -\p{{\mathcal F}_0}{\bar z_i} + \mathcal U_i\;, \label{active_nematic_defect dynamics}\eeq
 and the only difference from the passive case is an important new term
	\beq
	{\mathcal U}_i = \int d^2z \sqrt{g}\bar\partial_i \alpha_0 \mathcal I \label{eq:activeForce},
	\eeq
	which accounts for the contribution of the activity-induced velocity field, where
	\beq \mathcal I = -(v \partial + \bar v \bar\partial) \alpha_0 - i(\partial v - \bar \partial \bar v) + i (p/2 - 1)(v\partial\varphi - \bar v \bar\partial \varphi).\label{eq:I} \eeq
    
    We emphasize that Equations~\eqref{active_nematic_defect dynamics}-\eqref{eq:I} present a general description of the dynamics of active topological defects for any $p$-atic texture on an arbitrary geometry characterized by the metric $g_{z\bar{z}}=e^{\varphi}/2$. Within this general description, the defect phase $\alpha_0$, the free energy $\mathcal{F}_0$, and the velocity field $v$ are, respectively
    \begin{align}
     \alpha_0 &= -\frac{i}{2}p\sigma_i \ln\left(\frac{z - z_i}{\bar z - \overline{z_i}}\right),\\
     \mathcal F_0 &= -\frac{\pi p^2}{2} (K+K') \left(\sigma_i - \frac{1}{2}\sigma_i^2\right)\varphi(z_i),\\
     v &= \frac{1}{\mu}\nabla \cdot \sigma^{\text{active}}.
    \end{align}

\subsection{Active stress for a nematic}\label{sec:activenem}

    While for a general $p$-atic, multiple contributions to the active stress are allowed by virtue of the $p$-fold rotational symmetry of the order parameter~\cite{giomi2022hydrodynamic}, the most well-established form of the active stress is the stresslet contribution, which has been extensively observed in biological systems~\cite{doostmohammadi2018active,doostmohammadi2021physics} and studied theoretically. The stresslet contribution leads to dipolar active forces and is proportional to the nematic ($p=2$) tensor
    \beq \sigma^{\text{active}} = \zeta_Q Q^{zz},\label{eq:activestress} \eeq
    where $\zeta_Q$ is the scalar activity coefficient that characterizes the strength of the stresslet (and corresponding force dipole), and $Q^{zz}$ is the nematic tensor, i.e. a special case of the generic rank $p$ tensor $T$ in Sec.~\ref{sec:formulation} for $p=2$. As such, the nematic tensor in isothermal coordinates $Q^{zz}$ can be expressed in terms of its more familiar Cartesian components through Eq.~\eqref{Qzz}. Therefore, using the definition of the active stress in terms of the stresslet contribution (Eq.~\eqref{eq:activestress}), the velocity field in the overdamped limit (Eq.~\eqref{eq:v}) can be written in terms of the nematic tensor in isothermal coordinates as
     \beq v^z = \frac{\zeta_Q}{\mu}\nabla_z Q^{zz} = \zeta \nabla_z Q^{zz} = \zeta \left(\partial\varphi + i\partial\alpha\right)Q^{zz},\nonumber\label{eq:viso} \eeq
where $\zeta = \zeta_Q/\mu$. From this point onward, for ease of notation, we will suppress $z$ indices. Thus we write the activity-induced velocity field as $v = \zeta \nabla Q$ using this simplified notation.\\

\noindent{\bf Active nematic defects in flat geometry.}
	In order to apply the above framework to active nematics, we first review and rederive the well-known motile force for a $+1/2$ defect in flat space, where the geometric potential $\varphi=0$. In complex coordinates $z$ and $\bar z$, the nematic tensor ${\bf T} = {\bf Q}$ for a defect of charge $\sigma_i = \pm 1/2$ at the origin takes the form
	\beq Q = e^{i\alpha} = \left(\frac{z - z_i}{\bar z - \overline{z_i}}\right)^{\pm 1/2}. \eeq
	
	\begin{figure}[t]
		\centering
        \subfloat[]{\includegraphics[width=.45\columnwidth]{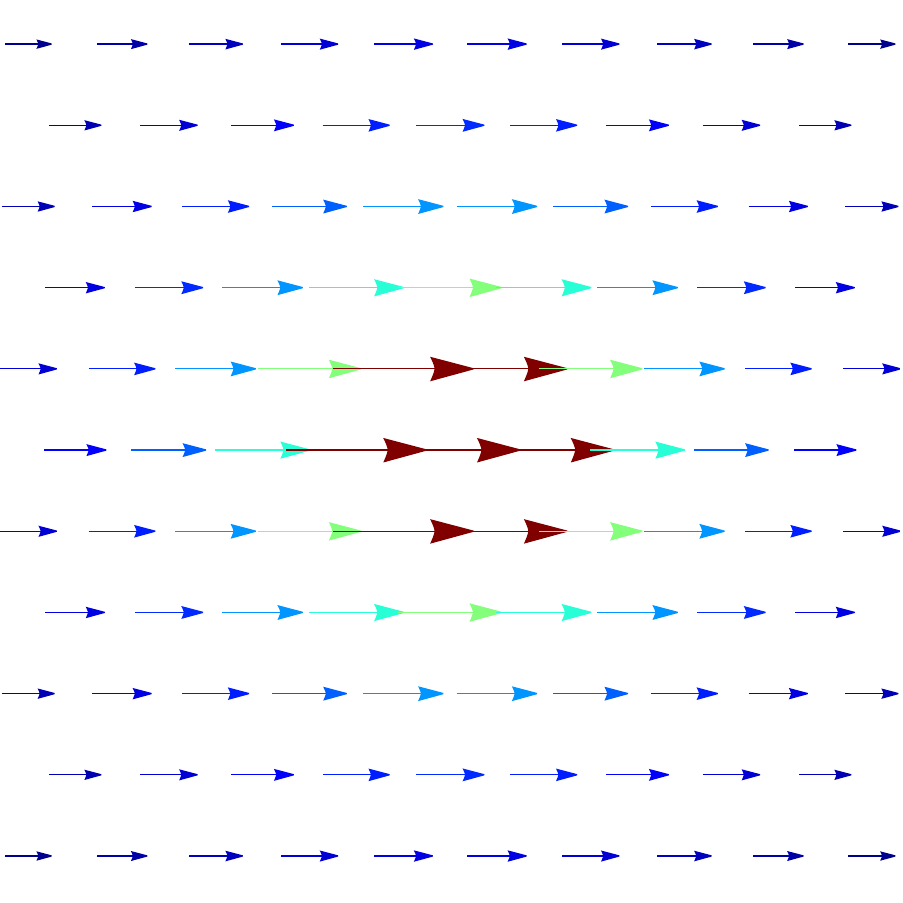}}
        \hfill
        \subfloat[]{\includegraphics[width=.45\columnwidth]{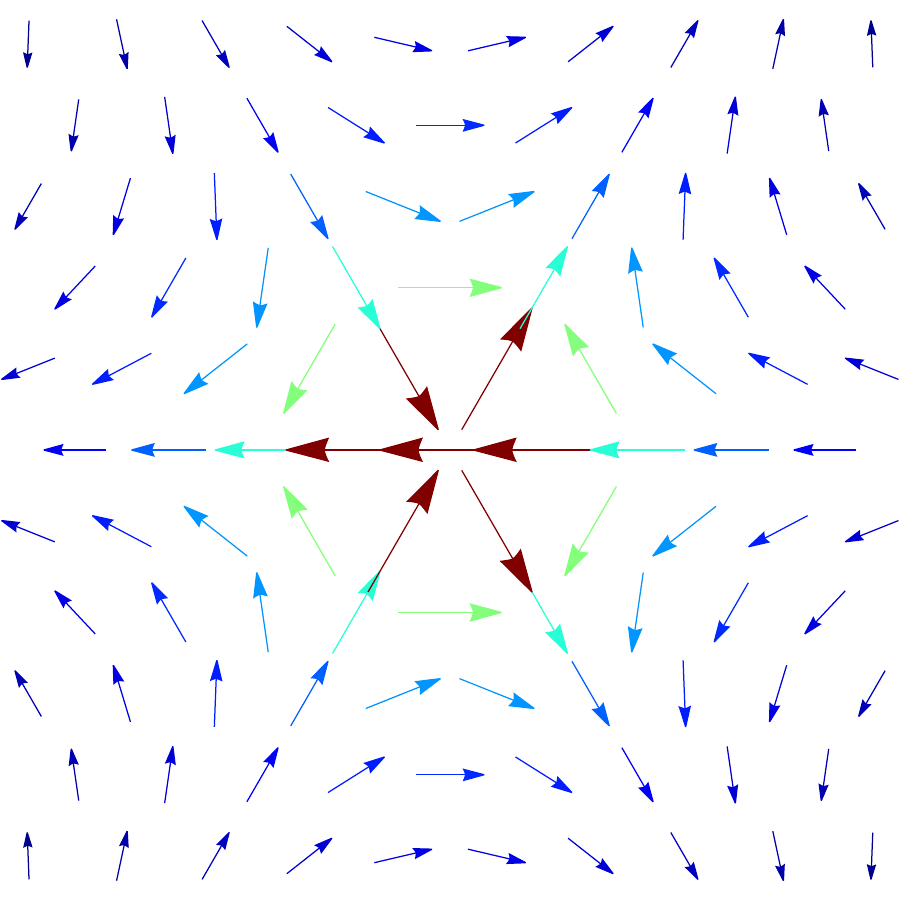}}
		\caption{Velocity profile of half-integer defects in flat space for $\zeta > 0$. (a): Velocity field generated by a $+1/2$ defect. (b): Velocity field generated a by $-1/2$ defect. The colormap represents velocity magnitude normalized by its maximum value, thus ranging from $0$ (dark blue) to $1$ (dark red).}
		\label{fig:v}
	\end{figure}
	Fig.~\ref{fig:v} presents the vector field corresponding to the velocity profile $\vec v = \zeta\nabla \cdot Q$ induced by contractile active stresses ($\zeta > 0$) around $\pm 1/2$ topological defects
    \begin{subequations}
	\begin{align}
		v^{+1/2} &= \frac{1}{2}\frac{1}{|z - z_i|}, \\
		v^{-1/2} &= -\frac{1}{2}\frac{|z - z_i|}{(z - z_i)^2}.
	\end{align}
    \end{subequations}
	Consistent with the theoretical~\cite{giomi2014defect}, numerical~\cite{doostmohammadi2018active}, and experimental~\cite{meacock2021bacteria} velocity fields for active nematic defects on a flat space, the flow field for a comet-shaped $+1/2$ defect points along the head-tail axis of the comet. However (unless external stresses or boundary conditions break the 3-fold rotational symmetry), on average any net flow for a $-1/2$ vanishes. These results are consistent with the intuition that a $+1/2$ defect, which has an asymmetric comet shape, moves along the comet axis (with a direction depending on the sign of $\zeta$), whereas a $-1/2$ defect, which has three-fold rotational symmetry, doesn't move by symmetry.
~\\

\noindent{\bf Active nematic defects on a cone.}
	We can now generalize the dynamics of active nematic defects to curved space. As before, $v = \zeta \nabla Q$, where the sign of $\zeta$ dictates contractile vs. extensile activity; in particular,
	\beq v = \zeta \left(\partial \varphi + i \partial\alpha\right)Q . \eeq
	Here we will focus on the case of a cone, in which case
	\beq \varphi = -\chi \ln z \bar z, \eeq
	where $2\pi\chi$ is the deficit angle (see Fig.~\ref{fig:cone}). We are as usual working in isothermal coordinates $z$ and $\bar z$.
	
	Upon focusing on the case of a single $+1/2$ nematic defect, we have
	\begin{align} 
    v &= \zeta\left(-\chi \frac{1}{z} + \frac{1}{2}\frac{1}{z - z_i}\right)Q \nonumber \\
    &= \zeta e^{-\varphi} \left(-\chi \frac{1}{z} + \frac{1}{2}\frac{1}{z - z_i}\right) \left(\frac{z - z_i}{\bar z - \overline{z_i}}\right)^{1/2},\nonumber\label{eq:vz}
    \end{align}
where $\varphi \neq 0$ reflects the fact that the metric is no longer flat.

\begin{figure}[t]
		\centering
		\subfloat[]{\includegraphics[width=.45\columnwidth]{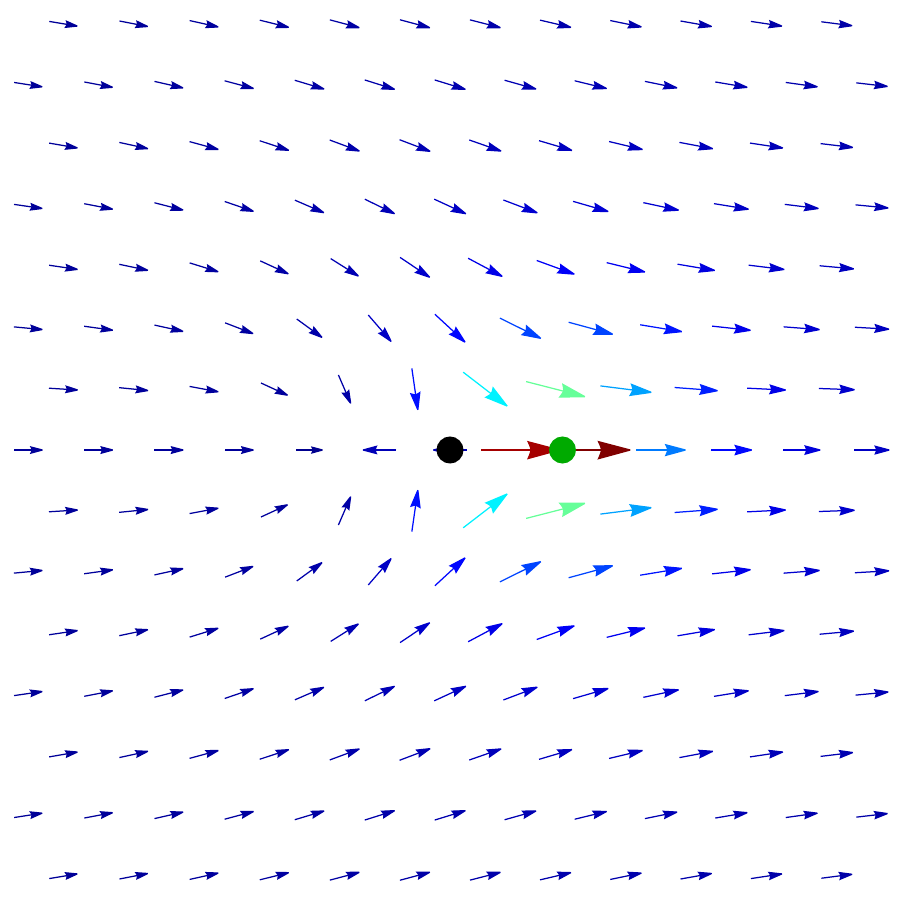}}
  \hfill
		\subfloat[]{\includegraphics[width=.45\columnwidth]{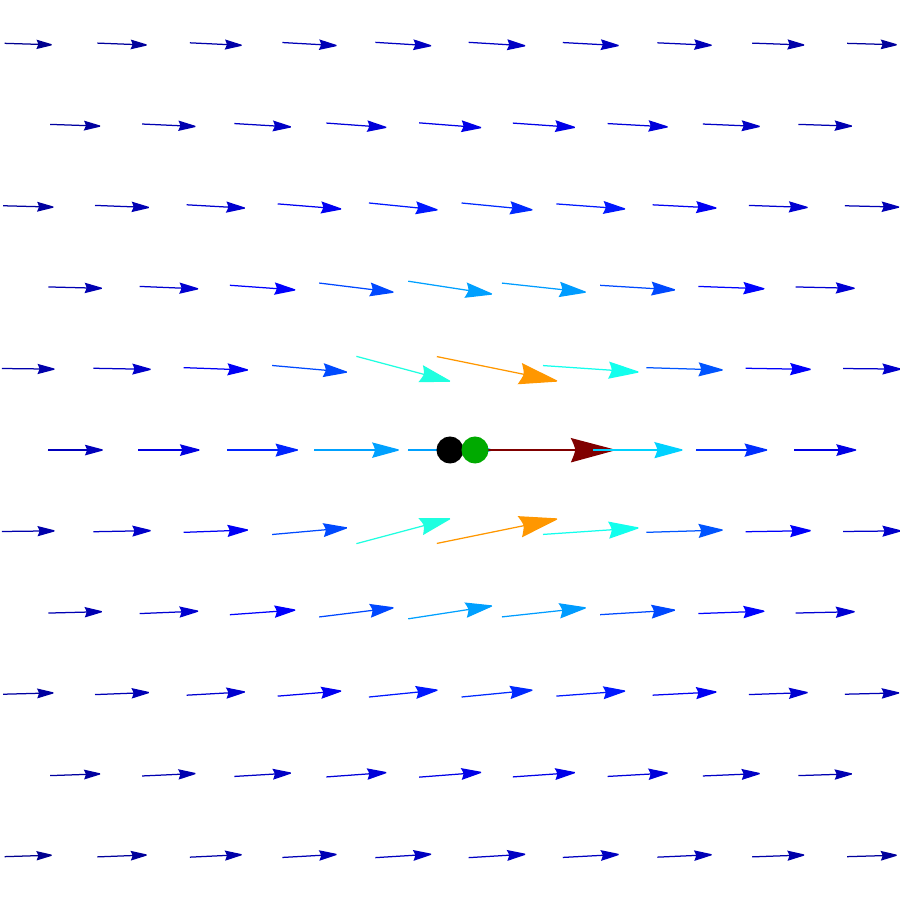}}\\
		\subfloat[]{\includegraphics[width=.45\columnwidth]{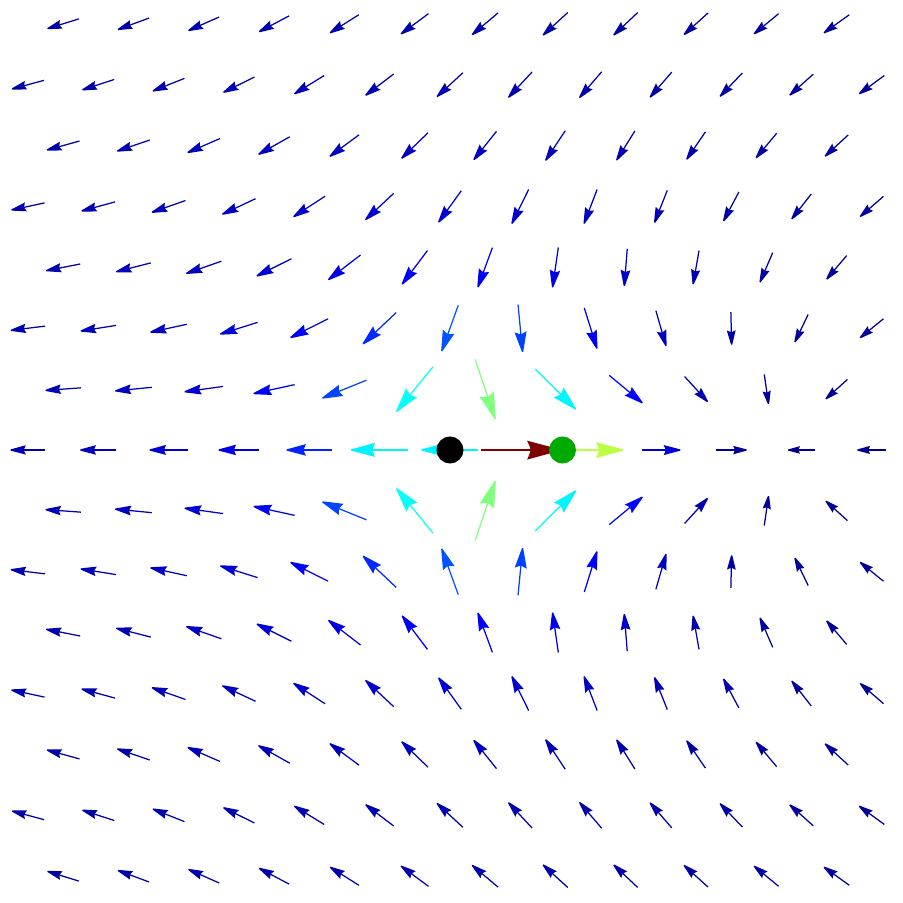}}
  \hfill
		\subfloat[]{\includegraphics[width=.45\columnwidth]{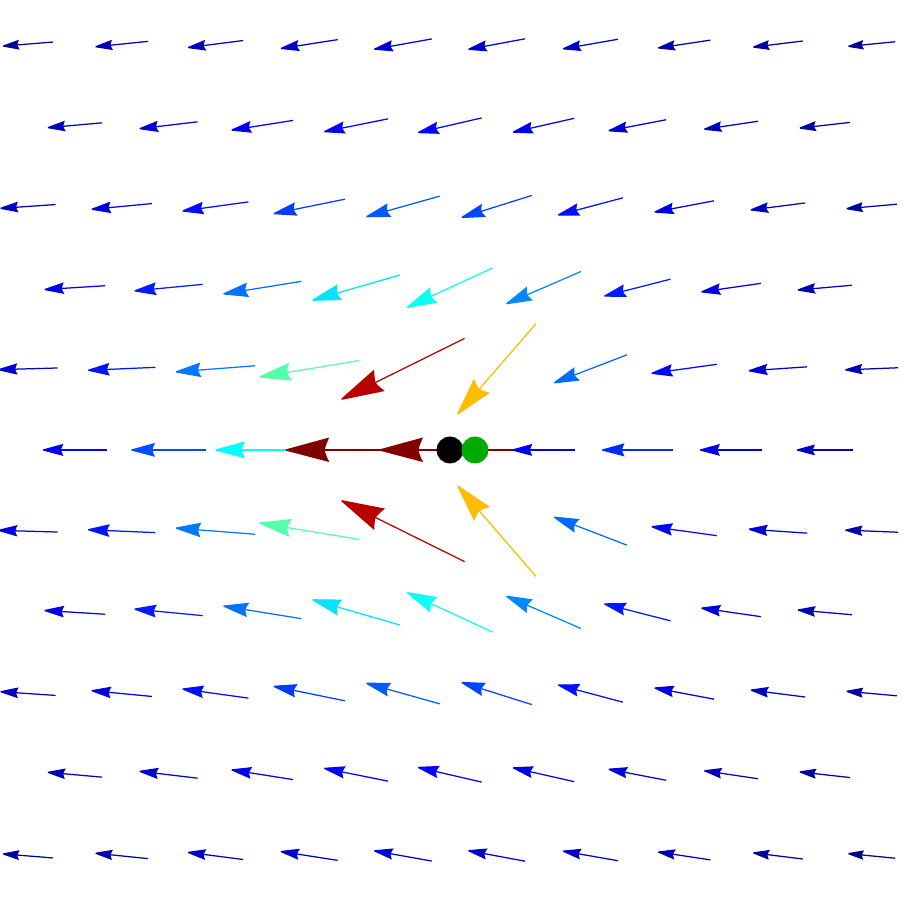}}
		\caption{Velocity profile produced by a contractile ($\zeta > 0$) $+1/2$ defect on a cone in the isothermal coordinates of Fig.~\ref{fig:cone} for a $+1/2$ defect at $(1,0)$ for $\chi = 0.25$ (top row) and $\chi = 0.75$ (bottom row). The black dot represents the apex and the green dot represents the $+1/2$ defect. Note that the velocity profile near the apex is similar for both $\chi = 0.25$ and $\chi=0.75$ (left column), but far away (right column, zoomed out) the velocity profile flips sign for $\chi = 0.75$. The colormap represents velocity magnitude normalized by its maximum value, thus ranging from $0$ (dark blue) to $1$ (dark red).}
		\label{fig:motility}
	\end{figure}

We comment that in the far-field, which is equivalent to taking a defect at the apex, i.e. $z_i \to 0$, $v^z$ reduces to
	\beq v \approx \zeta e^{-\varphi}\left(\frac{1}{2}-\chi\right)\frac{1}{z} \left(\frac{z}{\bar z}\right)^{1/2} = \zeta e^{-\varphi}\left(\frac{1}{2}-\chi\right)\frac{1}{|z|}, \label{eq:vzFar}\eeq
	which reveals the slow falloff of the activity induced velocity field and the special nature of cones with $\chi = 1/2$, i.e. cones formed from disks with exactly half their area removed.
	Thus $v$ changes sign as $\chi$ goes from below $1/2$ to above $1/2$, and the entire nematic texture moves accordingly along the real axis of our isothermal coordinate system. The vanishing velocity for $\chi = 1/2$ is reminiscent of (but not the same as) the passive case, where the apex charge is completely screened for $\chi = 1/2$. See Fig.~\ref{fig:motility} for plots of the velocity profile for a $+1/2$ defect for $\chi = 0.25$ and $\chi = 0.75$. Although the velocity fields generated by activity are similar near the cone apex, they are quite different in the far field.

\subsection{Evaluation of dynamical equation}

With the activity-induced velocity-field discussed, the active contribution to the dynamics of nematic defects can be calculated using Eq.~\eqref{eq:activeForce}, where setting $p=2$ and $\sigma= +1/2$ for active nematics leads to (Appendix~\ref{app:activeForcing})
	\beq I = \frac{\pi}{4}\frac{r_i^{-\chi}}{a}(1-\chi)(1-2\chi),\eeq
where $a$ is a short distance cutoff we take to be the spacing between liquid crystal molecules. Note that $I$ vanishes for $\chi = 1$ (corresponding to a long cylindrical shell) and $\chi = 1/2$.

We now combine the active velocity contribution, together with the calculations of the collective mobilities on the cone in Sec.~\ref{sec:passive} (Eqs.~\eqref{eq:Mcone1}-\eqref{eq:Mcone2}), and the equations of motion (Eq.~\eqref{active_nematic_defect dynamics}) for an active nematic $+1/2$ topological defect on a cone geometry to obtain our final equation for $+1/2$ defect dynamics on a cone,
	\begin{empheq}[box=\fbox]{align}
    &r_i^{-\chi}\ln(r_i^{(1-\chi)}/a) \dot z_i \nonumber\\
            & \qquad{} = \zeta   \frac{1}{a}(1-\chi)(1-2\chi) -3\gamma^{-1} \chi (K+K') \frac{r_i^{\chi}}{\overline{z_i}} .\label{eq:dzidtmain}
    \end{empheq}
	In terms of the physical coordinates shown in Fig.~\ref{fig:cone}(b),
	\beq \tilde z_i = \tilde r_i e^{i\tilde \phi_i} = \frac{z_i^{1-\chi}}{1 - \chi},\label{eq:main}\eeq
we have
\begin{align}
    	\ln[(1-\chi)\tilde r_i/a]\frac{d}{dt}\tilde z_i  = & \zeta e^{-i\frac{\chi}{1 - \chi}\tilde \phi_i}  \frac{1}{a}(1-\chi)(1-2\chi) \nonumber\\
        &-3\gamma^{-1} (K+K') \frac{\chi}{1-\chi} \frac{1}{\overline{\tilde z_i}}.
        \label{eq:dzitildedtmain}
\end{align}
 	
	It is easy to interpret the different contributions to the defect dynamics displayed in Eq.~\eqref{eq:dzidtmain} (or equivalently, Eq.~\eqref{eq:dzitildedtmain}): The LHS is the mobility, with a logarithmic correction that depends on the distance to the apex, unchanged by the activity. The first term on the RHS is the motile force caused the activity parameter $\zeta$, and the second term on the RHS is the attractive Coulomb force that a $+1/2$ defect feels from the cone apex with effective topological charge $-\chi$. The geometric corrections to the mobility, active force, and the interaction force are manifest through the dependence on the deficit angle $2\pi\chi$ of the cone. In particular, the motility term reverses sign for $\chi > 1/2$, consistent with the velocity profile in Eq.~\eqref{eq:vzFar}. In addition, the motility speed is reduced by a factor of $1-\chi$. Note that the angle of self-propulsion depends on the position, and, remarkably, for a critical defect position $\tilde z_i$, a $+1/2$ defect can have a stationary radius, which occurs when the RHS vanishes. This is where the Coulomb and motile forces balance each other and (as discussed further below) defines a basin of attraction for the positive defects.
	
	The radius of the basin of attraction (when the magnitude of the RHS vanishes), is
	\beq \tilde r_c = 3\zeta^{-1}\gamma^{-1} (K+K') \frac{\chi}{(1-\chi)^2|1-2\chi|} a. \label{eq:r_ctilde}\eeq
	Note that the critical radius $\tilde r_c$ diverges as the activity $\zeta$ tends to zero. Interestingly, for finite activity $\zeta$, the defect is always absorbed by the apex in the limit $\chi\to 1$, which is the limiting geometry of a cylinder, regarded here as an extremely pointed cylinder.

	\subsection{Defect trajectories}
	
    Having derived the equation of motion for the nematic defects in Eq.~\eqref{eq:dzidtmain} (or equivalently, Eq.~\eqref{eq:dzitildedtmain}), we then calculate defect trajectories. To summarize, the dynamical equations for a $+1/2$ defect in isothermal coordinates $z$ and physical coordinates $\tilde z$, read, respectively,
 	\begin{align}
		r^{-\chi}\ln(r^{(1-\chi)}/a) \dot z &= A -B \frac{r^{\chi}}{\overline{z}}, \label{eq:dzdt}\\
		\ln[(1-\chi)\tilde r/a]\frac{d}{dt}\tilde z  &= A e^{-i\frac{\chi}{1 - \chi}\tilde \phi} -\frac{B}{1-\chi} \frac{1}{\overline{\tilde z}} ,
	\end{align}
	where for a particular defect $i$ we define
	\begin{align}
		A &= \zeta   \frac{1}{a}(1-\chi)(1-2\chi), \\
		B &= 3\gamma^{-1} \chi (K+K'), \label{eq:B}\\
		z &= z_i, \\
		\tilde z &= \tilde z_i,\\
            r &= |z_i|.
	\end{align}

	Dividing Eq.~\eqref{eq:dzdt} by its complex conjugate results in
	\beq \frac{dz}{d\bar z} = \frac{A -B \frac{r^{\chi}}{\overline{z}}}{A -B \frac{r^{\chi}}{z}} . \label{eq:dzdzbar}\eeq

	\begin{figure}[t]
		\centering
		\subfloat[]{\includegraphics[width=.49\columnwidth]{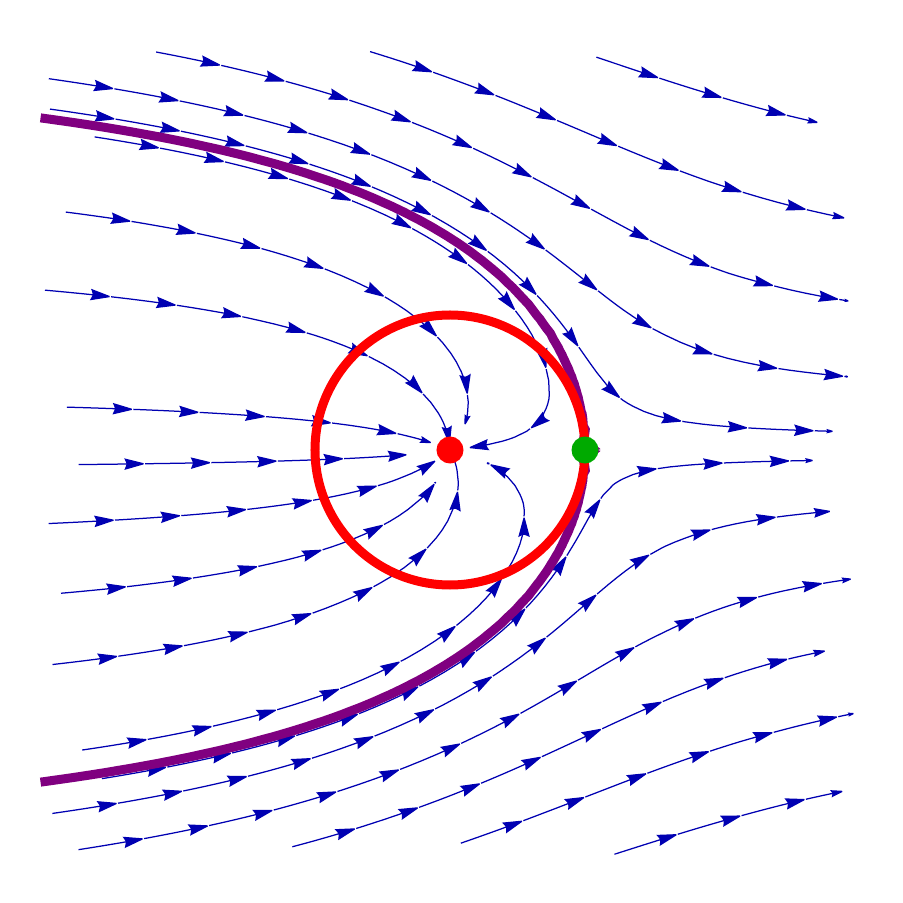}}
		\subfloat[]{\includegraphics[width=.49\columnwidth]{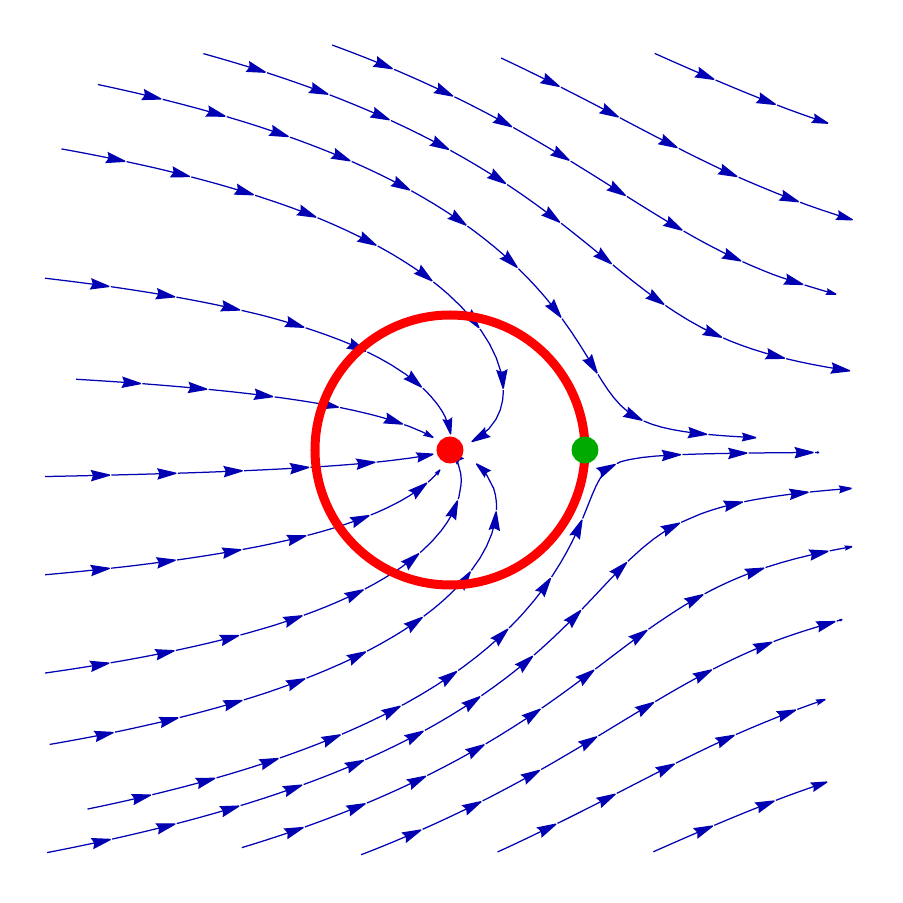}}\\
  		\subfloat[]{\raisebox{1cm}%
                    {\includegraphics[width=.49\columnwidth]{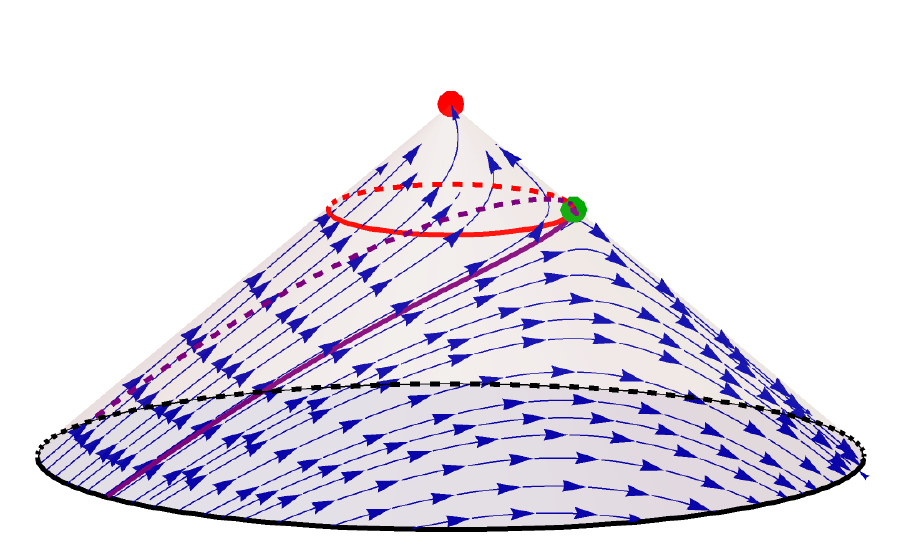}}}
		\subfloat[]{\includegraphics[width=.49\columnwidth]{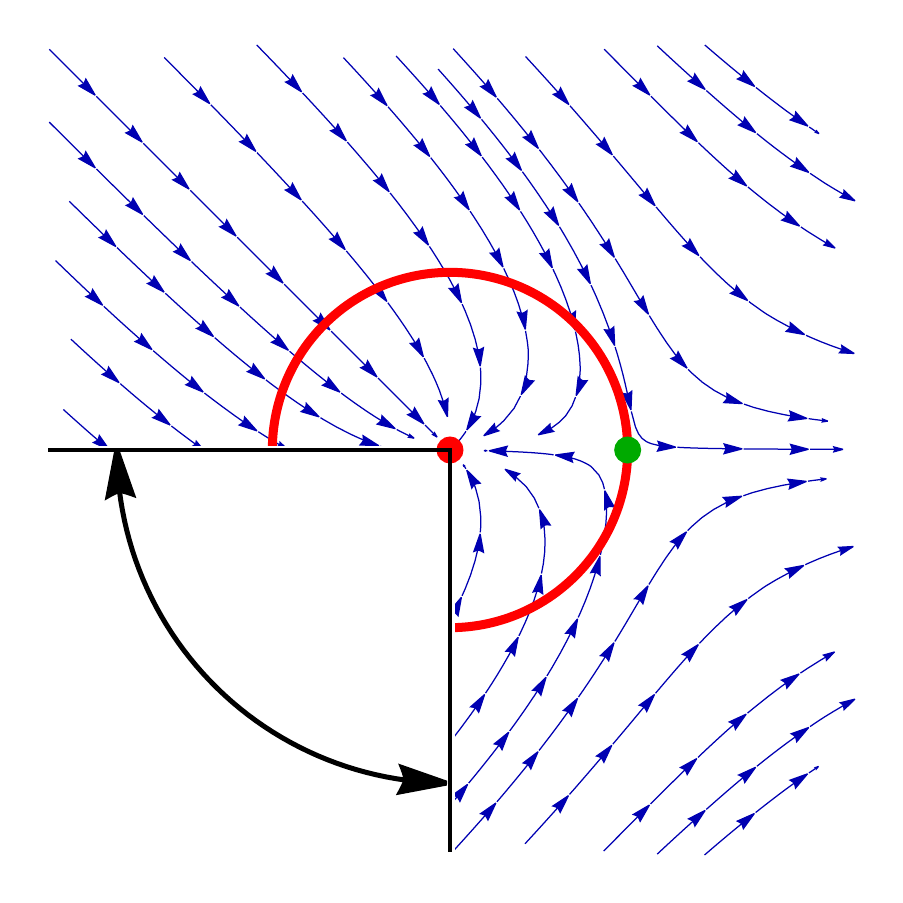}}
		\caption{Streamlines of defect trajectories for small $\chi=1/4$. Top row: streamlines in isothermal coordinates $z$, where (a): small $\chi$ analytic trajectories, given by Eq.~\eqref{eq:trajectory} and (b): numerical trajectories obtained by numerically integrating the exact ODEs, given by Eq.~\eqref{eq:dzdzbar}. (c): trajectories in physical coordinates wrapped around a 3D cone. (d): trajectories on an unrolled cone in $\tilde z$ coordinates where black arrows indicate identification of the two black edges. In all of the plots, the red dot denotes the cone apex, and the green point is where the motile and Coulomb forces on the $+1/2$ defect balance each other. Any point within a distance given by the radius of the red circle given by $\tilde r_c$ in Eq.~\eqref{eq:r_ctilde} gets attracted to the apex, and the magenta curve is an incoming separatrix. We expect that trajectories outside of the apex domain of attraction go down the cone flanks and are eventually influenced by the boundary conditions at the cone base. Other parameters: $A=B=a=1$.}
		\label{fig:contours}
	\end{figure}

	For $\chi = 0$, we recover the known result of $dz = dx + idy = d \bar z$, i.e., a defect with $dy = 0$ moving parallel to the real axis, which we assume is aligned with the ``comet-tail'' of the $+1/2$ defect. We now determine dynamical trajectories to leading order in $\chi$. From Eq.~\eqref{eq:B}, $B = \mathcal O(\chi)$, leading to
	\beq \frac{dz}{d \bar z} = \frac{A - \frac{B}{\bar z} + \mathcal O(\chi^2)}{A - \frac{B}{z} + \mathcal O(\chi^2)} .\eeq
 Upon integrating after cross-multiplying (and ignoring the $\mathcal O(\chi)$ corrections to the above equation), we have
	\beq A (z - z_0) - B \ln \frac{z}{z_0} = A (\bar z - \overline{z_0}) - B \ln \frac{\bar z}{\overline{z_0}}, \eeq
	where $z_0$ is the initial position of the defect.
	 Redefining our coordinates as $w = \frac{A}{B} z$ and rearranging leads to
	\beq (w - \bar w) - \ln \frac{w}{\bar w} = (w_0 - \overline{w_0}) - \ln \frac{w_0}{\overline{w_0}}. \eeq
 In terms of $w = w_R + i w_I = r e^{i\phi}$, we arrive at
	\beq w_I(t) - w_I(t=0) = \phi(t) - \phi(t=0) \label{eq:trajectory},\eeq
	where each set of initial conditions $(w_I(t=0),\phi(t=0))$ gives a different trajectory. Upon remembering that 
 \begin{align}
 w = \frac{A}{B}z &= \frac{\zeta \gamma}{3a(K+K')}\frac{(1-\chi)(1-2\chi)}{\chi }z \nonumber \\
 &= \frac{\zeta \gamma}{3a(K+K')}\frac{1}{\chi }z\left(1 + \mathcal O(\chi^2)\right),
 \end{align}
 we can compute defect trajectories, up to corrections of $\mathcal O(\chi^2)$. Fig.~\ref{fig:contours} illustrates the contours of defect trajectories around the cone apex, showing that the defect can get attracted to or repelled from the apex depending on the original position. 
 
\subsection{Validity of the Born-Oppenheimer approximation for $\chi \ge 1/2$}

It is important to note that for $\chi \ge 1/2$, since the motility for the global defect texture reverses sign, then it is as if the global texture is moving in the opposite direction of the local defect position. In other words, the Born-Oppenheimer ansatz in the near-field of the defect is not a good approximation for $\chi \ge 1/2$. It is possible that the Born-Oppenheimer approach could nevertheless be useful for understanding global properties of the texture, far away from the defect.

   \subsection{Comparison with simulations}

When activity is included, as a check on our analytic results, we numerically evolve the full nematic texture $Q$ according to
\beq D_t Q = \partial_t Q + (v\nabla + \bar v \bar\nabla) Q - (\nabla v - \bar\nabla \bar v)Q = 0 ,\eeq
where $\partial_t Q$ is given in Eq.~\eqref{eq:num} and $v = \zeta \nabla Q$. \\

\begin{figure}[t]
		\centering
		\includegraphics[width=\columnwidth]{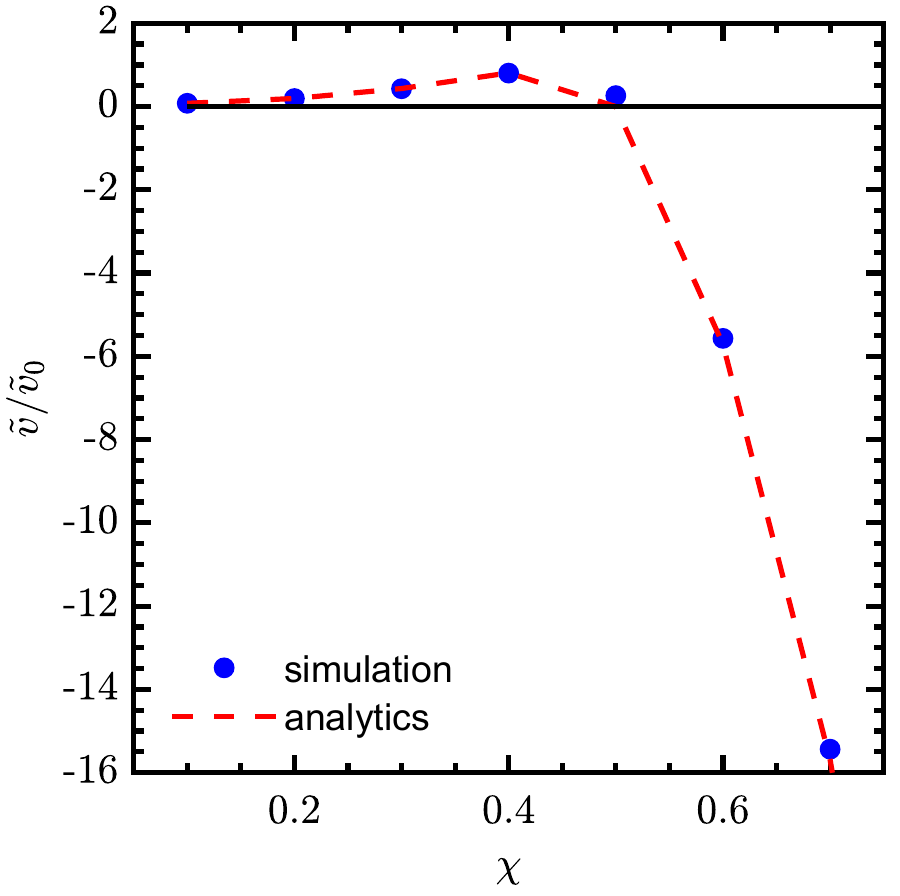}
		\caption{Initial physical global defect velocity due to motility for a $+1/2$ defect initially near the apex ($r(t=0) \ll R$). Blue points are obtained by fitting the texture at each time to the ansatz and determining $z_i(t)$ which minimize the mean squared deviation $E$ in Eq.~\eqref{nematic_eq:E}. The red curve is fit of Eq.~\eqref{eq:dzidtNear} from motile term with $\delta/R = 0.04$ and the size of region of optimization is $R'/R = 0.08$. The $y$-axis is rescaled by the characteristic active velocity $\tilde v_0 = \zeta /\tilde R$ where $\tilde R$ is the physical radius of the base of the cone.}
		\label{fig:dzidt}
	\end{figure}	
 
     \noindent{\bf Sign of global defect velocity.} Simulations indeed confirm that the local defect velocity never changes direction as $\chi$ is varied, as predicted by dynamical equations such as Eq.~\eqref{eq:dzidtmain} (or equivalently, Eq.~\eqref{eq:dzitildedtmain}). From simulations, we can also extract the global defect position by fitting the defect ansatz to the global texture data. When the defect is near the apex, i.e. $r_i\sim \delta$, the global defect velocity due to motility (i.e., ignoring the Coulomb term) is obtained by replacing the mobility in Eq.~\eqref{eq:dzidtmain} with Eq.~\eqref{eq:Mcone2}, leading to the global defect velocity,
    \beq \dot z_i = 2\zeta\frac{1}{\delta} \frac{1}{\left(\delta^{-2\chi} - R^{-2\chi} \right)}\chi(1-\chi)(1-2\chi)\label{eq:dzidtNear},\eeq
	where we used $\delta = a r_i^{\chi}$ in the above equation. The global velocity does indeed change sign as $\chi$ increases from below $1/2$ to above $1/2$. See Fig.~\ref{fig:dzidt} for comparison of simulations with theoretical prediction of global defect velocity due to the motility term of Eq.~\eqref{eq:dzidtNear}. The fact that the motility flips sign indicates that the behavior of extensile and contractile activity flips at $\chi = 1/2$, i.e., when the integral of the total Gaussian curvature associated with the cone apex is $\pi$.	\\

	\begin{figure}[t]
		\centering
        \subfloat[]{\includegraphics[width=.49\columnwidth]{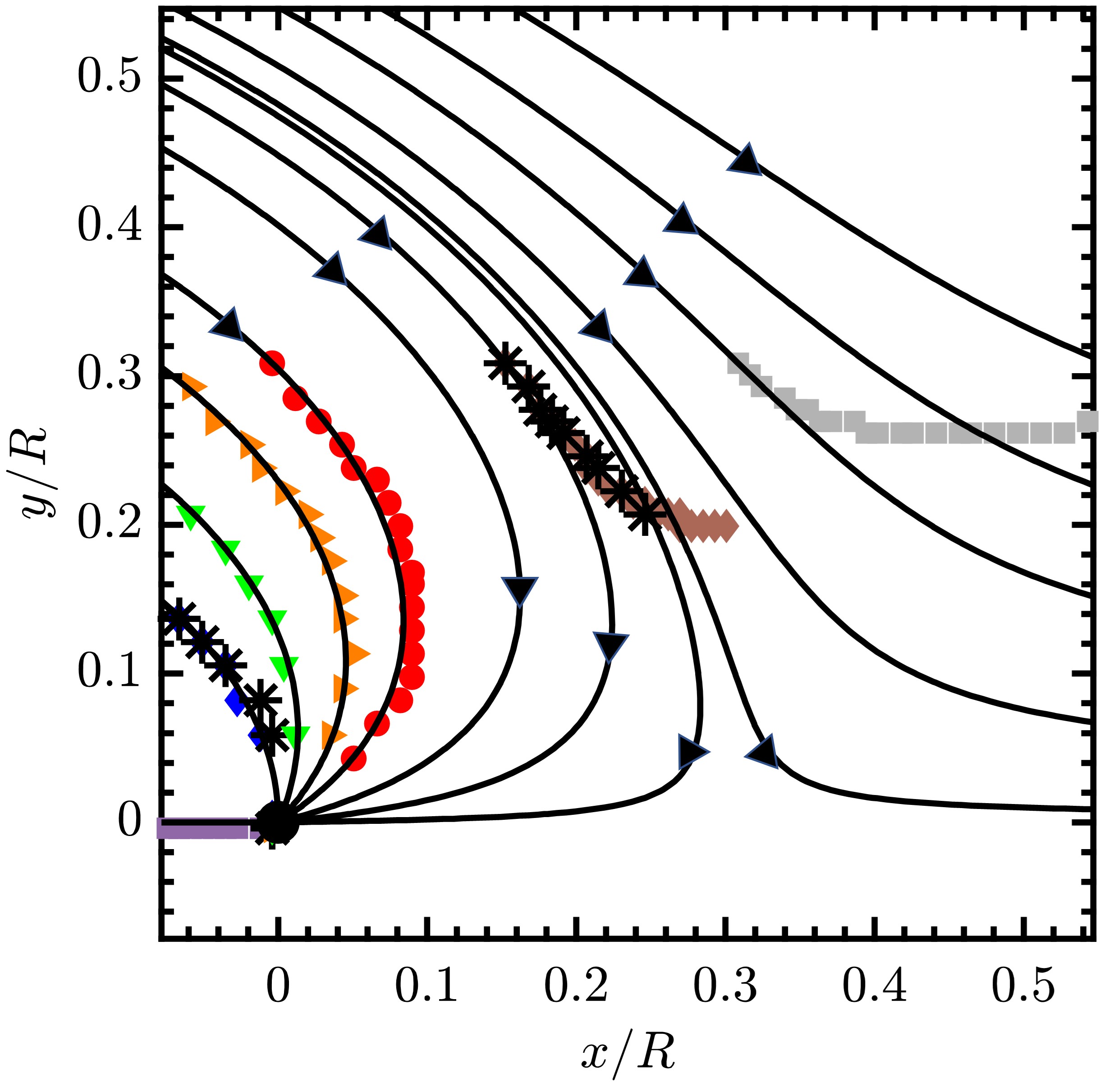}}
        \subfloat[]{\includegraphics[width=.49\columnwidth]{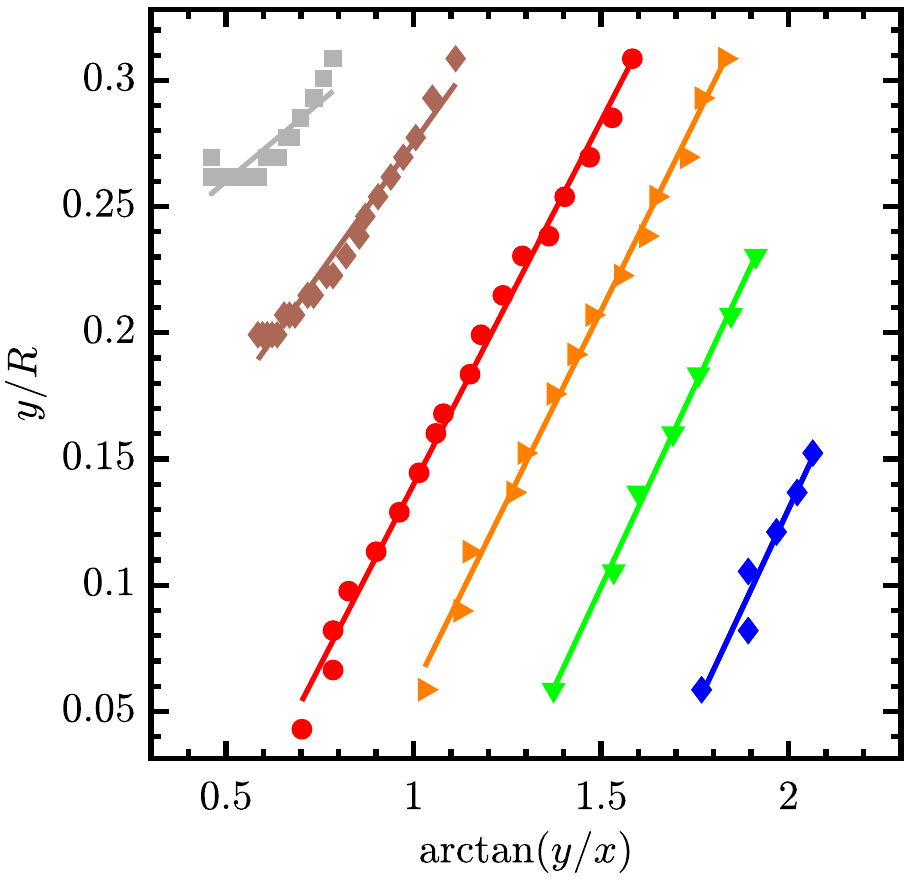}}
		\caption{Left plot: various defect trajectories using isothermal coordinates obtained from simulations and our analytical approximation. Colored markers are the local defect positions determined by simulation (each one corresponding to a different initial position), where the black asterisk markers denote simulations that used free boundary conditions for comparison. The black curves are predicted contours (Eq.~\eqref{eq:trajectory}) from the analytic approximation of the global defect position. Trajectories on the left side are attracted to the cone apex, while those on the flanks run away toward the base, where they will eventually be influenced by the boundary conditions at the cone base. The cone apex is at the origin (denoted by the black dot). The right plot contains linear best fits for each trajectory. Simulation parameters: $\chi = 0.1$ and $\zeta/[\gamma^{-1}(K+K')] = 2\times 10^{-2}$, and $a/(\sqrt{K+K'}\epsilon) =0.67$. Radius of the disk in $z$ coordinates is $R=128$, but since this size is large enough, it does not affect the trajectories near the apex.}
		\label{fig:trajectories}
	\end{figure}
 
 \noindent{\bf Defect trajectories.} For small $\chi$ ($\chi = 0.1$) and small activity ($\zeta/[\gamma^{-1}(K+K')] = 2\times 10^{-2}$), we used simulations to evolve the nematic texture consisting of a defect initially near the apex. As before, we extract the global defect position by fitting the defect ansatz to the global texture data and plot in Fig.~\ref{fig:trajectories}. Eq.~\eqref{eq:trajectory} predicts that all of the slopes are the same for global defect positions. Comparison to numerical simulations confirms this prediction for defects close to the cone apex (Fig.~\ref{fig:trajectories}; {\it blue, green, orange, and red lines}), with the slope $m \approx  0.3 R$. On setting $B/A = m$ and solving for $a$, we find $a/(\sqrt{K+K'}\epsilon) = 0.67$. Simulation results further show that the prediction for the slope becomes worse as the initial distance of the defect to the apex increases (Fig.~\ref{fig:trajectories}; {\it brown and grey lines}), which is consistent with the fact that as time passes we expect the ansatz to become less accurate.

	\section{Discussion}
    \label{sec:discussion}

With the help of the Born-Oppenheimer approximation, which assumes liquid crystal textures can relax instantaneously to defect positions, we have presented a description of topological defects dynamics in compressible active nematic materials deep in the ordered phase on curved surfaces with a focus on the behavior near the apex of cones. We find that activity induces an active geometric contribution to the motility of the global texture of the $+1/2$ defect which changes sign as the curvature increases, a result we expect to hold more generally. Furthermore, in the case of a cone, we provide a closed form prediction for the radius of the basin of attraction around the apex, and present analytical description of the defect trajectories near the apex. Sufficiently far from the cone base, the analytical results agree well with full numerical simulations.

We would like to emphasize that although our analysis focused on the interaction between an active nematic defect and the cone apex, the formulation presented here is general and can be applied to geometries with more general conformal factors and $p$-atic textures. Furthermore, since we assumed that the defect was near the apex and far away from the boundary of the cone, we could safely ignore the effect of boundary conditions. The framework can be extended to study defect interactions with the boundaries through introducing image charges, which is the focus of our upcoming study; preliminary investigations have revealed a rich phase diagram of allowed dynamical states, which exhibits not only single flank defect orbits and two flank defect orbits, but also transitions between them via defect absorption, defect emission, and defect pair creation via activity at the apex~\cite{vafa2023}. We expect significantly different results for defect dynamics closer to the boundary, with, say, tangential as opposed to free boundary conditions.

In a related direction, it would also be worth exploring defect configurations on cones with both tangential and free boundary conditions at finite temperatures. With increasing temperatures, entropic effects might cause the cone apex to emit defects or induce defect unbinding.

Finally, we note that the formulation presented here could be a useful tool for studying topological defects interacting on non-trivially curved surfaces that biologically active entities experience in various setups. One striking example is the interaction of topological defects with dynamic protrusions that are formed from cellular membranes~\cite{leijnse2022filopodia}, cell monolayers~\cite{guillamat2022integer}, or the cell cytoskeleton~\cite{maroudas2020topological}. Such protrusions are often characterized by non-trivial, dynamically changing, surface curvature, and even sharp tips, which according to our analyses can result in dynamic geometric contributions to the motility of active topological defects. We find in upcoming work~\cite{vafa2023} that activity can catalyze sharp, lightning-rod-like conical geometries to create and emit defects. 

Furthermore, there is currently intense research in designing in-vitro surfaces for tissue regeneration~\cite{matejvcic2022mechanobiological}. Topological defects are also finding increasing applications in governing biological functions such as cell death and removal, and cell differentiation. In this vein, our predictive framework can potentially find applications in designing surface geometries that allow for prescribed locations of topological defects.

\acknowledgments{It is a pleasure to acknowledge helpful conversations with Siavash Monfared, Suraj Shankar, and Grace Zhang. This work is partially supported by the Center for Mathematical Sciences and Applications at Harvard University (F. V.), and by the Harvard Materials Research Science and Engineering Center via Grant DMR-2011754 (D.R.N.).  F.V. and A.D. gratefully acknowledge Nordita for their hospitality, where some of the research for this paper was performed during the 2022 Summer workshop ``Current and Future Themes in Soft \& Biological Active Matter''. F.V. additionally gratefully acknowledges the Niels Bohr Institute for their hospitality, where some of the research for this paper was performed. A. D. acknowledges funding from the Novo Nordisk Foundation (grant No. NNF18SA0035142 and NERD grant No. NNF21OC0068687), Villum Fonden (Grant no. 29476), and the European Union (ERC, PhysCoMeT, 101041418). Views and opinions expressed are however those of the authors only and do not necessarily reflect those of the European Union or the European Research Council. Neither the European Union nor the granting authority can be held responsible for them.}

\bibliography{refs}

\newpage
  \onecolumngrid
	\section*{Appendix}

	\appendix
 
	\section{Derivation of multi-defect dynamics equation}
	\label{app:derivation}
	
	To describe nematic dynamics in the limit of weak activity and low defect density, we shall assume that the phase field $\alpha(z,\bar z,t)$ of the order parameter texture stays close to the inertial manifold $\alpha_0 (z, {\bar z}|\{ z_i (t)\})$ parameterized by time-dependent defect positions:
	\beq
	\alpha(z,{\bar z}, t) = \alpha_0 (z, {\bar z}|\{ z_i (t)\})+\delta \alpha(z,{\bar z}, t)\;,
	\eeq
	where $\delta \alpha$ is locally perpendicular to the inertial manifold as defined by
	\begin{align}
		\int d^2z \sqrt{g} \ \partial_i \alpha_0 \delta\alpha = 
		\int d^2z \sqrt{g} \ {\bar \partial}_i \alpha_0 \delta \alpha = 0\;. \label{eq:orthogonality}
	\end{align}
	We thus rewrite the complex texture dynamics equation Eq.~\eqref{eq:complexQ} as
	\begin{align}
		{\dot z}_j \partial_j \alpha_0+{\dot {\bar z}}_j{\bar \partial}_j \alpha_0  &+\partial_t \delta \alpha
		=\mathcal I \nonumber\\
		&=-\frac{1}{\sqrt{g}}\frac{\delta {\mathcal F}[\alpha]}{\delta \alpha}
		+\zeta \mathcal I_{\zeta}[\alpha]\;.
	\end{align}
	Multiplying by $\bar\partial_i \alpha_0$ and integrating over space, we find that
	\beq \mathcal M_{ij}\dot z_j  + \mathcal N_{ij}\dot{\bar z}_j   = \int d^2z \sqrt{g} \ \bar\partial_i \alpha_0 \mathcal I\;, \label{nematic_defectEqn}\eeq
	where
	\begin{align}
		\mathcal M_{ij} &= \int d^2z \sqrt{g} \ \bar\partial_i \alpha_0 \partial_j \alpha_0\\
		\mathcal N_{ij} &= \int d^2z \sqrt{g} \ \bar\partial_i \alpha_0 \bar \partial_j \alpha_0\;.
	\end{align}
	Up to now, the discussion has been general. We will now work in the limit of small activity $\zeta \ll 1$ and large defect separation $\epsilon^{-1} \gg 1$. In this limit,  $\delta \alpha \ll \alpha_0$ because the phase field $\alpha_0 (z,{\bar z}|\{ z_i \})$ minimizes the LdG free energy to order ${\mathcal O} (\epsilon^2)$ on the punctured plane with fixed $z_i$. Thus to leading order
	\beq \mathcal I(\alpha) \approx \mathcal I(\alpha_0)\;.\eeq
	Now using the fact that
	\beq \p{\mathcal F}{\bar z_i} = \int d^2z \sqrt{g} \  \bar\partial_i\alpha_0 \frac{\delta \mathcal F}{\delta \alpha_0}\;,
	\label{nematic_eq:Coul}\eeq
	we find that
	\begin{align}
		\int d^2z \sqrt{g} \ \bar\partial_i \alpha_0 \mathcal I &= \int d^2z \sqrt{g} \  \bar\partial_i \alpha_0 [-\frac{\delta {\mathcal F}}{ \delta \alpha_0}+\zeta \mathcal I_\zeta(\alpha_0)] \nonumber\\
		& = -\p{\mathcal F}{\bar z_i} + \zeta\int d^2z \sqrt{g} \ \bar\partial_i \alpha_0 \mathcal I_\zeta \;.
	\end{align}
	To summarize, our defect dynamics equations are
	\beq \mathcal M_{ij}\dot z_j  + \mathcal N_{ij}\dot{\bar z}_j   = -\p{{\mathcal F}_0}{\bar z_i} + {\mathcal U}_i \;,\eeq
	with
	\beq
	{\mathcal U}_i = \zeta\int d^2z \sqrt{g} \ \bar\partial_i\bar \alpha_0 \mathcal I_\zeta\label{nematic_eq:active force}
	\eeq
	the active forcing.
	
	It is perhaps not surprising that the equations of motion for $z_i(t)$ that we have obtained minimize the mean squared deviation of the dynamics on the inertial manifold $Q_0$ from the exact equation of motion Eq.~\eqref{eq:complexQ}. That is, we minimize
	\begin{align}
		E &= \int d^2z \sqrt{g}\left| \partial_t \alpha( z,{\bar z},t) - \frac{d}{dt} \alpha_0( z,{\bar z}|\{ z_i (t) \})\right|^2 \nonumber\\
		&\approx \int d^2z \sqrt{g} \left| \mathcal I [\alpha_0] - \dot z_i\partial_i \alpha_0 - \dot{\bar z}_i \bar\partial_i \alpha_0\right|^2
	\end{align}
	with respect to $\dot z_i$.

 	\section{Computation of collective mobilities}
    \label{app:mobilities}

  Using Eqs.~\eqref{nematic_eq:M}-\eqref{nematic_eq:N}, the collective mobilities are calculated as
	
	\begin{align}
		\mathcal M_{ij} &= \frac{1}{8}(p\sigma_i)(p\sigma_j)\int d^2z (z\bar z)^{-\chi}\frac{1}{\bar z - \overline{z_i}}\frac{1}{z - z_j} \\
		\mathcal N_{ij} &=\frac{1}{8}(p\sigma_i)(p\sigma_j)\int d^2z (z\bar z)^{-\chi}\frac{1}{\bar z - \overline{z_i}}\frac{1}{\bar z - \overline{z_j}}.
	\end{align}
	In particular, for a single defect
	\begin{align}
		\mathcal M_{ii} &= \frac{1}{8}(p\sigma_i)^2 \int d^2z (z\bar z)^{-\chi}\frac{1}{|z - z_i|^2}, \\
		\mathcal N_{ii} &= \frac{1}{8}(p\sigma_i)^2  \int d^2z (z\bar z)^{-\chi}\frac{1}{(\bar z - \overline{z_i})^2}
	\end{align}
	Since $\mathcal N_{ii}$ is finite, and $\mathcal M_{ii}$ diverges near $z=z_i$,  we will ignore the subleading $\mathcal N_{ii}$ term and focus on the dominant $\mathcal M_{ii}$ contribution, which can be exactly evaluated in terms of incomplete beta functions.

    First we evaluate $\mathcal M_{ii}$ assuming the defect is sufficiently far from the apex, i.e., $r_i = |z_i| \gg a^{1/(1-\chi)}$. Rescaling $z \to z_i z$ leads to
 	\beq \mathcal M_{ii} = \frac{1}{8}(p\sigma_i)^2 r_i^{-2\chi} \int d^2z (z\bar z)^{-\chi}\frac{1}{|z - 1|^2},\label{eq:Mii}\eeq
 We now split the integral in Eq.~\eqref{eq:Mii} into two regions: (i) $r = |z| < 1$ and $r > 1$,  and then matching powers of $z$ with powers of $\bar z$. This leads to
	\beq \mathcal M_{ii} = \frac{\pi}{8} (p\sigma_i)^2 r_i^{-2\chi} \left[\int_{r=0}^{r=1 - \delta/r_i} d(r^2) \frac{r^{-2\chi}}{1 - r^2} + \int_{r=1 + \delta/r_i}^{r=\infty} d(r^2) \frac{r^{-2\chi}}{r^2 - 1}\right].\label{eq:Mii-far} \eeq
	We thus have to properly regularize the integral by choosing appropriate $\delta$. The integral is formally infinite, but this ignores the defect core size $a$, which sets a natural UV cut-off. By  definition of the metric, the distance 
	\beq 
	d(z_m,z_n) = e^{\varphi(z_m)/2}|z_m - z_n| = a 
	\eeq
	for small separation the order of the defect core size, and thus
	\beq \delta = |z_m - z_n| = a e^{-\varphi(z_m)/2}.
	\eeq
	Thus, in isothermal coordinates, the defect core cut-off $\delta$ for defect at $z_i$ is taken to be $\delta = a e^{-\varphi(z_i)/2}$, which for a cone ($\varphi = -\chi \ln z\bar{z}$) is evaluated to be
	\beq \delta = a r_i^\chi \eeq
	
	Using this cut-off length the dominant contributions to the mobility of a single defect, Eq.~\eqref{eq:Mii-far} are
	\begin{align}
		\int_0^{1 - \delta/r_i} d(r^2) \frac{r^{-2\chi}}{1 - r^2} &\approx -\ln (\delta/r_i) \\
		\int_{1 + \delta/r_i}^\infty d(r^2) \frac{r^{-2\chi}}{r^2 - 1}  &\approx - \ln (\delta/r_i)
	\end{align}
	and thus
	\beq \mathcal M_{ii} = \frac{\pi}{4} p^2\sigma_i^2 r_i^{-2\chi}\ln(r_i^{(1-\chi)}/a),  \eeq
	which describes the mobility of a single $p$-atic topological defect of charge $\sigma_i]$, with the core radius $a$, located far from the apex of a cone with the cone deficit angle of $2\pi\chi$.
	When the defect is close to the apex, instead we find
	\beq \mathcal M_{ii} = \frac{1}{8}(p\sigma_i)^2 \int d^2z r^{-2\chi}\frac{1}{r^2} =  \frac{\pi}{4}(p\sigma_i)^2\frac{1}{2\chi} \left(\delta^{-2\chi} - R^{-2\chi} \right) \label{eq:mobilityNear}\eeq
	where $\delta = a r_i^{-\chi} \sim a^{1/(1-\chi)}$ for $r_i \sim \delta$.\\

    \section{Computation of active forcing}
    \label{app:activeForcing}

    Here we explicitly compute $\mathcal U_i$ (Eq.~\eqref{eq:I}). Upon substituting
	\begin{align}
		\nabla Q_0 &= \left(\partial\varphi + i\partial\alpha_0\right)Q_0\\
		\bar\nabla Q_0 &= \left(-\bar\partial\varphi + i\bar\partial\alpha_0\right)Q_0 \\
		v &= \zeta\nabla Q_0 \\
		\partial v &= [\partial^2\varphi + i\partial^2\alpha_0]Q_0 - [(\partial\varphi)^2 + (\partial\alpha_0)^2]Q_0
	\end{align}
	in Eq.~\eqref{eq:I} we have
	\begin{align}
		\mathcal I_\zeta &=  - \partial \alpha_0 (\partial\varphi + i\partial\alpha_0) Q_0 - \bar\partial \alpha_0 (\bar\partial\varphi - i\bar\partial\alpha_0) \bar Q_0 \nonumber \\
		& \quad - i[\partial^2\varphi + i\partial^2\alpha_0]Q_0 + i[(\partial\varphi)^2 + (\partial\alpha_0)^2]Q_0 \nonumber\\
		& \quad + i[\bar\partial^2\varphi - i\bar\partial^2\alpha_0]\bar Q_0 - i[(\bar\partial\varphi)^2 + (\bar\partial\alpha_0)^2]\bar Q_0 \nonumber \\
		&=  - \partial\alpha_0 \partial\varphi Q_0 - \bar\partial \alpha_0 \bar\partial\varphi \bar Q_0  - i[\partial^2\varphi - (\partial\varphi)^2 + i\partial^2\alpha_0]Q_0 + i[\bar\partial^2\varphi - (\bar\partial\varphi)^2 - i\bar\partial^2\alpha_0]\bar Q_0 \label{eq:Iactive}
	\end{align}
from which the active contribution to the defect dynamics (Eq.~\eqref{active_nematic_defect dynamics}) can be calculated 
	\beq \mathcal U_i = \int d^2z \sqrt{g}\bar\partial_i \alpha_0 \mathcal I_\zeta = I_1 + I_2\eeq
	where
	\begin{align}
		I_1 &= \frac{1}{2}\int d^2z e^{i\alpha_0} \bar\partial_i \alpha_0 \left[\partial^2\alpha_0 -\partial\alpha_0\partial\varphi - i\partial^2\varphi + i (\partial\varphi)^2\right] \\
		I_2 &= \frac{1}{2}\int d^2z e^{-i\alpha_0} \bar\partial_i \alpha_0 \left[\bar\partial^2\alpha_0 -\bar\partial\alpha_0\bar\partial\varphi + i\bar\partial^2\varphi - i (\bar\partial\varphi)^2\right]
	\end{align}
	Substituting
	\beq \bar\partial_i \alpha_0 = -\frac{i}{2}\frac{1}{\bar z - \overline{z_i}}\eeq
	gives
	\begin{align}
		I_1 &= \frac{1}{4}\int d^2z \frac{(z - z_i)^{1/2}}{(\bar z - \overline{z_i})^{1/2}} \frac{1}{\bar z - \overline{z_i}} \left[-i\partial^2\alpha_0 +i\partial\alpha_0\partial\varphi - \partial^2\varphi + (\partial\varphi)^2\right] \\
		I_2 &= \frac{1}{4}\int d^2z \frac{(\bar z - \overline{z_i})^{1/2}}{(z - z_i)^{1/2}}  \frac{1}{\bar z - \overline{z_i}} \left[-i\bar\partial^2\alpha_0 + i\bar\partial\alpha_0\bar\partial\varphi + \bar\partial^2\varphi - (\bar\partial\varphi)^2\right]
	\end{align}
	
	We first note that in $I_1$ and $I_2$, the $\partial^2\alpha_0$ and $\bar\partial^2\alpha_0$ terms were previously found in \cite{vafa2020multi-defect}. In particular, the $\partial^2\alpha_0$ term in $I_1$ gives rise to the motility of the $+1/2$ defect, and the $\bar\partial^2\alpha_0$ term in $I_2$ gives rise to the pair-wise defect interactions. In both integrals, the terms involving $\varphi$ are the active contributions of the force from the geometry, which for small activity are subleading (compared to the usual Coulomb force). Before we start computing, we note that because of phase symmetry, $I_2$ vanishes. It thus suffices to compute $I_1$.
	
	We begin by computing the first term in $I_1$. We have
	
	\beq \int d^2z \frac{(z - z_i)^{1/2}}{(\bar z - \overline{z_i})^{1/2}} \frac{1}{\bar z - \overline{z_i}} (-i)\partial^2\alpha_0 = \frac{1}{2}\int d^2z \frac{(z - z_i)^{1/2}}{(\bar z - \overline{z_i})^{1/2}} \frac{1}{\bar z - \overline{z_i}} \frac{1}{(z - z_i)^2} = \frac{1}{2}\int d^2z \frac{1}{|z - z_i|^3}\eeq
	Upon shifting $z \to z + z_i$,
	\beq\int d^2z \frac{1}{|z - z_i|^3} = \int d^2z \frac{1}{r^3} =2\pi \int_\delta^\infty dr \frac{r}{r^3} = 2\pi \frac{1}{\delta}  =2\pi \frac{r_i^{-\chi}}{a}\eeq
 
	We continue to compute.
	
	\beq \int d^2z \frac{(z - z_i)^{1/2}}{(\bar z - \overline{z_i})^{1/2}} \frac{1}{\bar z - \overline{z_i}} (i)\partial\alpha_0\partial\varphi =  -\frac{1}{2}\chi  \int d^2z \frac{(z - z_i)^{1/2}}{(\bar z - \overline{z_i})^{1/2}} \frac{1}{\bar z - \overline{z_i}}\frac{1}{z - z_i}\frac{1}{z}\eeq
	
	Upon shifting $z \to z + z_i$,
	\beq  \int d^2z \frac{(z - z_i)^{1/2}}{(\bar z - \overline{z_i})^{1/2}} \frac{1}{\bar z - \overline{z_i}}\frac{1}{z - z_i}\frac{1}{z} =  \int d^2z \frac{1}{r^2} \frac{z^{1/2}}{\bar z^{1/2}}\frac{1}{z+z_i} =2\pi \int_\delta^\infty dr \frac{r}{r^3} =2\pi \frac{r_i^{-\chi}}{a}\eeq
	
	We continue to compute.
	
	\beq \int d^2z \frac{(z - z_i)^{1/2}}{(\bar z - \overline{z_i})^{1/2}} \frac{1}{\bar z - \overline{z_i}} \left[- \partial^2\varphi + (\partial\varphi)^2\right] = -\chi(1-\chi)\int d^2z \frac{(z - z_i)^{1/2}}{(\bar z - \overline{z_i})^{1/2}} \frac{1}{\bar z - \overline{z_i}} \frac{1}{z^2}\eeq

	Upon shifting $z \to z + z_i$,
	\beq \int d^2z \frac{(z - z_i)^{1/2}}{(\bar z - \overline{z_i})^{1/2}} \frac{1}{\bar z - \overline{z_i}} \frac{1}{z^2} =  \int d^2z \frac{z^{1/2}}{\bar z^{1/2}} \frac{1}{\bar z} \frac{1}{(z+z_i)^2} =  \int d^2z \frac{1}{r^2} \frac{z^{1/2}}{\bar z^{1/2}}\frac{1}{z+z_i} =2\pi \int_\delta^\infty dr \frac{r}{r^3} =2\pi \frac{r_i^{-\chi}}{a}\eeq
	Thus,
	\beq I_1 = \frac{\pi}{4}\frac{r_i^{-\chi}}{a}\left(1 - \chi - 2\chi(1-\chi)\right) = \frac{\pi}{4}\frac{r_i^{-\chi}}{a}(1-\chi)(1-2\chi)\eeq
	
	It is interesting that for all $\chi$, the motility is reduced, and in fact reverses sign for $\chi > 1/2$. It is also interesting to note that for $\chi = 1/2$ and $\chi = 1$, $I_1$ vanishes, in which case a $+1/2$ defect is not motile.
	
	Combining everything,
	\beq \boxed{r_i^{-\chi}\ln(r_i^{(1-\chi)}/a) \dot z_i = \zeta   \frac{1}{a}(1-\chi)(1-2\chi) -3\gamma^{-1} \chi (K+K') \frac{r_i^{\chi}}{\overline{z_i}}\label{eq:dzidt}}\eeq
	In terms of physical coordinates
	\beq \tilde z_i = \frac{z_i^{1-\chi}}{1 - \chi},\eeq
	and using
	\beq  \dot z_i = [(1-\chi)\tilde z_i]^\frac{\chi}{1-\chi} \frac{d}{dt}\tilde z_i = z_i^\chi \frac{d}{dt}\tilde z_i\eeq
	we have
	\begin{align}
		&\ln \left[(1-\chi)\tilde r_i/a\right] e^{i\chi\phi} \frac{d}{dt}\tilde z_i =\zeta  \frac{1}{a}(1-\chi)(1-2\chi) -3\gamma^{-1} \chi (K+K') \frac{[(1-\chi)\tilde r_i]^{\frac{\chi}{1-\chi}}}{\left[(1 - \chi)\overline{\tilde z_i}\right]^\frac{1}{1 - \chi}}
	\end{align}
	resulting in
	\beq
	\ln[(1-\chi)\tilde r_i/a]\frac{d}{dt}\tilde z_i  = \zeta e^{-i\frac{\chi}{1 - \chi}\tilde \phi_i}  \frac{1}{a}(1-\chi)(1-2\chi) -3\gamma^{-1} (K+K') \frac{\chi}{1-\chi} \frac{1}{\overline{\tilde z_i}}
	\eeq

	\section{Time-dynamics of defect trajectory}
	\label{app:trajectory}
 
	We now calculate the time-dependence of the defect position for a defect trajectory along the $x$-axis. We have
	
	\beq x^{-\chi}\ln(x^{(1-\chi)}/a) \dot x = A -B x^{\chi-1}\eeq
	which is integrated to give trajectory $x(t)$ which satisfies
	\beq \int dx \frac{x^{-\chi}\ln(x^{(1-\chi)}/a)}{A -B x^{\chi-1}} = t\eeq
	\begin{align} 
		&t= \frac{x^{-\chi }}{2 A^2 (\chi -1)} \times \left(-2 \ln \left(\frac{x^{1-\chi }}{a}\right) \left(B x^{\chi } \ln \left(B-A x^{1-\chi }\right)+A x\right)+2 B x^{\chi } \text{Li}_2\left(\frac{B x^{\chi -1}}{A}\right) \right. \\
		& \left. -2 B (\chi -1) x^{\chi } \ln (x) \left(\ln \left(B-A x^{1-\chi }\right)-\ln \left(1-\frac{B x^{\chi -1}}{A}\right)\right)+2 A x-B (\chi -1)^2 x^{\chi } \ln ^2(x)\right)
	\end{align}
	where Li$_2$ is the dilogarithm.
	
\end{document}